\documentclass[preprint,aps,prd,superscriptaddress,nofootinbib,tightenlines]{revtex4}

\usepackage{amsfonts}
\usepackage{multirow}
\usepackage{mathrsfs}
\usepackage{graphicx}
\usepackage{amsmath}
\usepackage{amssymb}
\usepackage{bm}
\usepackage{bbm}
\usepackage{color}
\usepackage{xcolor}
\usepackage{float}
\usepackage{slashed}
\usepackage{ulem}

\newcommand{\nn}{\nonumber}
\newcommand{\beq}{\begin{equation}}
\newcommand{\eeq}{\end{equation}}
\newcommand{\bseq}{\begin{subequations}}
\newcommand{\eseq}{\end{subequations}}

\newcommand{\QCDtwo}{\text{QCD}_2}
\newcommand{\sQCDtwo}{\text{sQCD}_2}
\newcommand{\hQCDtwo}{\text{hQCD}_2}
\newcommand{\Nc}{N_\mathrm{c}}
\newcommand{\SUNc}{SU\!\left(\Nc\right)}
\newcommand{\gs}{g_\mathrm{s}}
\newcommand{\LF}{\text{LF}}
\newcommand{\pint}{\makebox[0pt][l]{$\mkern5mu-$}\int}

\usepackage{amsmath}
\usepackage{tikz}
\usepackage{adjustbox}
\usepackage{subcaption}

 \newcommand{\T}{\mathrm{T}}
 \newcommand{\tr}{\mathrm{tr}}
 \newcommand{\bra}[1]{\langle#1\vert}
 \newcommand{\ket}[1]{\vert#1\rangle}
 \newcommand{\specialcell}[2][c]{%
   \begin{tabular}[#1]{@{}c@{}}#2\end{tabular}}
 \newcommand{\loosebox}[1]{\adjustbox{padding = 1ex 1ex 1ex 1ex}{#1}}

\begin{document}

 \usetikzlibrary{shapes.misc}
 \usetikzlibrary{arrows.meta}
 \usetikzlibrary{calc}
 \usetikzlibrary{
  decorations.pathmorphing,
  decorations.pathreplacing,
  decorations.markings
 }
 \usetikzlibrary{patterns}
 \usetikzlibrary{shapes.symbols}
 \usetikzlibrary{shapes.geometric}

 \tikzset{
  every picture/.style={semithick, line cap=round},
  scalar/.style={dashed},
  fermion/.default=0.5,
  fermion/.style={postaction={decorate, decoration={
     markings,
     mark=at position #1 with {\arrow{Stealth[angle=30:7pt,inset=1.5pt]}},
     transform={xshift={3.5pt*cos(15)}}
  }}},
  antifermion/.default=0.5,
  antifermion/.style={postaction={decorate, decoration={
     markings,
     mark=at position #1 with {\arrowreversed{Stealth[angle=30:7pt,inset=1.5pt]}},
     transform={xshift={-3.5pt*cos(15)}}
  }}},
  gluon/.default=4pt,
  gluon/.style={decorate, decoration={
     coil,
     amplitude=0.5*#1,
     aspect=1,
     segment length=#1
  }},
  rgluon/.default=4pt,
  rgluon/.style={decorate, decoration={
     coil,
     amplitude=-0.5*#1,
     aspect=-1,
     segment length=#1
  }},
  gluonpre/.default=0pt,
  gluonpre/.style={decorate, decoration={
     coil,
     amplitude=2pt,
     aspect=1,
     segment length=4pt,
     pre length=#1
  }},
  rgluonpre/.default=0pt,
  rgluonpre/.style={decorate, decoration={
     coil,
     amplitude=-2pt,
     aspect=-1,
     segment length=4pt,
     pre length=#1
  }},
  wilson/.style={
     double distance=2pt,
     line cap=butt
  },
  dbline/.default=0.5,
  dbline/.style={postaction={decorate, decoration={
     markings,
     mark=at position #1 with {\arrow{Stealth[inset=0, length=5pt, angle'=45]}},
     transform={xshift={2.5pt}}
  }}},
  crossmark/.style={cross out, draw=red, inner sep=2pt},
  counter/.style={path picture={
     \draw (path picture bounding box.south east) --
      (path picture bounding box.north west)
      (path picture bounding box.south west) --
      (path picture bounding box.north east);
  }},
  cnode/.default=8pt,
  cnode/.style={inner sep=0pt, minimum size=#1, circle},
  trinode/.style={
     shape=isosceles triangle,
     fill=black, minimum width=8, inner xsep=0, xshift=-3,
     isosceles triangle apex angle = 60
  },
  rtrinode/.style={
     shape=isosceles triangle,
     fill=black, minimum width=8, inner xsep=0, xshift=3,
     isosceles triangle apex angle = 60,
     shape border rotate=180
  }
 }
\title{\mbox{}\\[10pt]
Solving bound-state equations in $\QCDtwo$ with bosonic and fermionic quarks}
\author{Xiaolin Li}
\affiliation{Department of Physics,Yantai University,Yantai 264005, China\vspace{0.2cm}}
\affiliation{Institute of High Energy Physics, Chinese Academy of
Sciences, Beijing 100049, China\vspace{0.2cm}}

\author{Yu Jia\footnote{Corresponding author: jiay@ihep.ac.cn}}
\affiliation{Institute of High Energy Physics, Chinese Academy of
Sciences, Beijing 100049, China\vspace{0.2cm}}
\affiliation{School of Physics, University of Chinese Academy of Sciences,
Beijing 100049, China\vspace{0.2cm}}

\author{Ying Li\footnote{Corresponding author: liying@ytu.edu.cn}}
\affiliation{Department of Physics,Yantai University,Yantai 264005, China\vspace{0.2cm}}

\author{Zhewen Mo\footnote{Corresponding author: mozw@itp.ac.cn}}
\affiliation{Institute of High Energy Physics, Chinese Academy of
Sciences, Beijing 100049, China\vspace{0.2cm}}
\affiliation{CAS Key Laboratory of Theoretical Physics, Institute of Theoretical Physics,
Chinese Academy of Sciences, Beijing 100190, China\vspace{0.2cm}}

\date{\today}
\begin{abstract}
We investigate the bound-state equations (BSEs) in two-dimensional QCD in the $\Nc\to \infty$ limit, viewed from both the infinite momentum frame (IMF)
and the finite momentum frame (FMF). The BSE of a meson in the original 't Hooft model, {\it viz.}, spinor $\QCDtwo$ containing only fermionc quarks,
has been extensively studied in literature. In this work, we focus on the BSEs pertaining to two types of ``exotic" hadrons, 
a ``tetraquark"  which is composed of a bosonic quark and bosonic antiquark, and a ``baryon"  which is composed of a bosonic antiquark and a fermionic quark.
Utilizing the Hamiltonian approach, we derive the corresponding BSEs for both types of ``exotic" hadrons,
from the perspectives of the light-front and equal-time quantization, and confirm the known results.
The recently available BSEs for ``tetraquark" in FMF has also been recovered with the aid of the diagrammatic approach.
For the first time we also present the BSEs  of a ``baryon" in FMF in the extended 't Hooft model.
By solving various BSEs numerically, we obtain the mass spectra pertaining to ``tetraquark" and ``baryon" and 
the corresponding bound-state wave functions of the lowest-lying states.
It is numerically demonstrated that, when a ``tetraquark" or ``baryon" is continuously boosted, the
forward-moving component of the bound-state wave function approaches the corresponding light-cone wave function,
while the backward-moving component fades away.
\end{abstract}

\maketitle

\section{Introduction}

Understanding the hadron structure from QCD is the central mission of the contemporary hadron physics.
Due to our limited knowledge about the color confinement mechanism, we are still unable to
write down, let alone to solve, the bound-state equations (BSEs) pertaining to any hadron in terms of the relativistic
quark and gluons degrees of freedom. There are some influential and powerful nonperturbative approaches,
such as Dyson-Schwinger (DS)/Bethe-Salpeter (BS) equations \cite{Roberts:1994dr, Bashir:2012fs, Chang:2009zb, Chang:2013pq},
as well as light-front (LF) quantization \cite{Brodsky:1997de, Li:2015zda, Lan:2019vui, Mondal:2019jdg},
which, under some approximation,
enable one to numerically solve the BSEs of hadron in Minkowski spacetime.
Unfortunately, at practical level, these approaches heavily depend on some unsystematized truncation, whose prediction is
thus subject to certain amount of model dependence.

The limit of infinite number of color, {\it viz.}, $1/\Nc$ expansion \cite{tHooft:1973alw, Witten:1979kh}, has proved to be a useful nonperturbative tool to
help us to understand a number of some essential phenomena of QCD,
such as Regge trajectory \cite{Witten:1979kh}, $U(1)_A$ problem \cite{Witten:1978bc, Veneziano:1979ec}, and so on.
Unfortunately, due to the enormous complexity of nonabelian gauge theory in four spacetime dimension,
at present we still do not know how to write down the BSEs for a hadron even in the large $N_c$ limit,
let alone to deduce the nonperturbative features of a hadron in a quantitative manner.

In 1974 't Hooft invented a solvable toy model of QCD, {\it i.e.}, QCD in two spacetime dimensions meanwhile in the $\Nc\to\infty$ limit~\cite{tHooft:1974pnl}.
Thanks to the absence of transverse degree of freedom of the gauge field together with the planarity of Feynman diagrams,
't Hooft was able to write down the bound state equation of a meson in a closed form. The resulting discrete mesonic energy levels
can be solved numerically, where the Regge trajectory becomes manifest.
Soon it was also realized that 't Hooft model has also possesses some interesting properties like (naive) asymptotic freedom \cite{Callan:1975ps},
nonvanishing quark condensate and ``spontaneous" chiral symmetry breaking \cite{Zhitnitsky:1985um, Burkardt:1995eb}.
Therefore, the 't Hooft model may be regarded as an instructive theoretical laboratory, which
may help us to gain some insight into the nonperturbative aspects of QCD in the real world
\cite{Callan:1975ps, Einhorn:1976uz, Einhorn:1976ev,Brower:1978wm, Li:1986gf, Li:1987hx,Kalashnikova:2001df,Jia:2017uul, Jia:2018qee}.

The BSE of 't Hooft model was originally derived with the aid of the diagrammatic technique
based on the DS/BS equations, also within the context of light-front quantization.
Therefore, the {\it 't Hooft equation} is valid only in the infinite momentum frame (IMF), {\it viz.},
in which a meson moves with the speed of the light. It is worth noting that,
an alternative approach to derive the 't Hooft equation is the operator approach based on bosonization of the
light-front Hamiltonian, which has been extensively discussed in
literature~\cite{Kikkawa:1980dc,Nakamura:1981zi,Dhar:1994ib,Dhar:1994aw,Cavicchi:1993jh,Barbon:1994au,Itakura:1996bk}.

Poincar\'{e} invariance demands that the meson mass spectra must be identical in any inertial reference frame.
One naturally wonders how the BSEs in 't Hooft model look like in the reference
frame other than IMF. An important progress was made by Bars and Green in 1978~\cite{Bars:1977ud}, who explicitly constructed a pair of coupled BSEs of mesons in
$\QCDtwo$ in the finite momentum frame (FMF), in which a meson moves with a finite momentum, including static case.
The resulting BSEs pertaining to FMF, dubbed {\it Bars-Green equation}, is much more involved than the 't Hooft equation pertaining to IMF.
It was also formally demonstrated that the Poincar\'{e} algebra does hold in the color-singlet subspace~\cite{Bars:1977ud}.
Later Poincar\'{e} invariance of meson spectrum has been explicitly verified by numerically solving Bars-Green in FMFs (static meson~\cite{Li:1986gf} and moving meson~\cite{Jia:2017uul}).

The advent of large momentum effective theory (LaMET)~\cite{Ji:2013dva,Ji:2014gla} allows one to
directly compute the partonic distributions of a hadron on the Euclidean lattice.
A key element of LaMET is that, the quasi distributions, which are the matrix element of the equal-time, spacelike correlator sandwiched between a moving hadron state,
under continuous Lorentz boost, will finally approach the light-cone parton distributions, which are the matrix element of the
light-like correlator sandwiched between a hadron in IMF.
 't Hooft model turns out to be a valuable theoretical laboratory to develop some intuition about the profiles of the quasi distributions.
In the 't Hooft model, the light-cone distribution is simply linked with the 't Hooft wave function,
and the quasi-distributions can be constructed in terms of the Bars-Green wave functions and Bogoliubov-chiral angle.
It has been analytically and numerically verified that, a variety of quasi distributions in the 't Hooft model does converge to the
light-cone distributions,
as anticipated from LaMET \cite{Jia:2018qee,Ma:2021yqx,Jia:2024atq}.

Shortly after 't Hooft's original work, Shei and Tsao~\cite{Shei:1977ci} in 1977 investigated the scalar $\QCDtwo$ which instead contains bosonic quarks.
With the aid of the diagrammatic approach in the context of LF quantization, Shei and Tsao also derived the BSE for a meson composed of a bosnic quark and a
bosonic antiquark. The {\it Shei-Tsao equation} looks very similar to the 't Hooft equation.
In 1978 Tomaras rederived the Shei-Tsao equation from the angle of Hamiltonian approach, and elaborated on some subtlety pertaining to quark mass
renormalization~\cite{Tomaras:1978nt}. Utilizing the Hamiltonian approach in the context of the equal-time
quantization, Ji, Liu and Zahed have recently also derived the BSEs in scalar $\QCDtwo$ pertaining to FMF~\cite{Ji:2018waw},
and showed that these BSEs do approach the Shei-Tsao equation when boosted to the IMF.

In the real world there is no bosonic quark. However, the notion of diquark turns out to be useful in baryon spectrum and structure, at least on the phenomenological ground ~\footnote{The diquark conjecture dates back to the early days of
the quark model~\cite{Ida:1966ev, Lichtenberg:1967zz}, by interpreting a baryon as a bound state composed of a diquark and another quark.
The predictions based on the diquark ansatz reduce the number of levels in the baryon spectrum, which get closer to the experiment data
than the three-body predictions in the quark model.  In order for the
lowest-lying octet and decuplet baryons to form a fully symmetric $\mathbf{56}$
representation of the spin-flavor $SU(6)$, the diquark in lowest-lying baryons
must be in a ${}^3S_1$ state, which corresponds to a spin-$1$ boson. Intuitively, a stable diquark may be formed since the gluonic exchange between two quarks in the $\bar{\mathbf{3}}$
representation of color $SU(3)$ is attractive.
Later some more sophisticated approaches~\cite{Cahill:1988dx, Reinhardt:1989rw, Efimov:1990uz},
{\it e.g.}, a Faddeev-type formulation which preserves the Lorentz and chiral symmetries, also adopt the diquark picture
Some recent development on diquark-based phenomenology can be found in Refs.~\cite{Chen:2017pse, Qin:2019hgk, Barabanov:2020jvn, Liu:2022nku}.}.
In 2008 Grinstein, Jora and Polosa investigated the mesonic mass spectra in scalar $\QCDtwo$~\cite{Grinstein:2008wm}, who argued that,
a bosonic quark may mimick a {\it diquark} to some extent, therefore the meson in scalar $\QCDtwo$ may be related to the tetraquark state in the real world,
which is conjectured to consist of a compact diquark and anti-diquark. It is expected that the study of the ``tetraquark" spectrum in scalar $\QCDtwo$~\cite{Grinstein:2008wm}
may shed some light on the tetraquark spectrum in the realistic QCD$_4$~\cite{Grinstein:2008wm}.

It is difficult to investigate the BSE for a baryon in the original 't Hooft model, since a baryon would become infinitely heavy in the $N_c\to \infty$ limit.
Nevertheless, once accepting the notion of the diquark, one may mimick a ``baryon" by a bound state formed by a bosonic quark and fermionic antiquark in the extended 't Hooft model.
The BSE for such a ``baryon" state in IMF was first obtained with the aid of diagrammatic technique by Aoki~\cite{Aoki:1993ma}.

It is the primary goal to derive the BSEs for a ``baryon" state in the extended 't Hooft model in FMF,
which constitute the counterparts of the Bars-Green equations of mesons in the original 't Hooft model.
We also conduct a comprehensive numerical study of the ``baryon" spectrum and the bound-state wave functions of the lowest-lying baryon.
It is especially interesting to visualize how the wave functions in the FMF evolve to the light-cone wave functions when increasing the momentum of the ``baryon".

We use the Hamiltonian approach in equal-time quantization with a ``fermionization" procedure to derive the BSEs of ``baryon" in FMF.
For the sake of completeness, we also revisit the derivation of BSE of ``baryon" in the IMF from the angle of the Hamiltonian approach in LF quantization.
Since the ``baryon" contains a bosonic quark, the BSEs of which are intimated connected with those of ``tetraquark" composed entirely of bosonic quark and antiquark.
Therefore, to facilitate a coherent reading, we feel it beneficial to give a self-contained treatment of both types of exotic hadrons.
Therefore we decide to revisit the derivations of the BSEs of the ``tetraquark in IMF and FMF using Hamiltonian approach,
which were originally done in \cite{Tomaras:1978nt,Ji:2018waw}.
Some subtle issue about the quark mass renormalization in LF and equal-time quantization in scalar $\QCDtwo$ is highlighted.
Moreover, it is worth pointing out that the diagrammatic derivation of the BSEs of a ``tetraquark" in FMF is much difficult  than its counterpart in IMF,
since the seagull vertex does not vanish even after taking the axial gauge.
A novel outcome of this work is to successfully reproduce the BSEs of ``tetraquark"
from the angle of the diagrammatic DS/BS approaches.

The rest of the paper is organized as follows.
In Sec.~\ref{setup:stage}, we define the extended 't Hooft model and set up some notations.
In Sec.~\ref{sec:scalar}, we revisit the
derivation of BSEs of an ``tetraquark" in both IMF and FMF, with some emphasis on the relation between the renormalized
quark mass in LF and equal-time quantization.
Sec.~\ref{sec:hybrid} constitutes the main body of this work, where we derive the BSEs of a ``baryon" in both IMF and FMF.
The BSEs in FMF are obtained for the first time.
In Sec~\ref{sec:numerical} we present the numerical results of the mass spectra of ``tetraquark" and ``baryon".
In particular, we show the numerical profiles of the bound-state wave functions of the lowest-lying states,
with different hadron velocities.
We summarize in Sec.~\ref{sec:summary}.
We devote Appendix~\ref{Diagram:derivation:sQCD2} to a diagrammatic derivation of the corresponding BSEs of ``tetraquark" in FMF.
In Appendix~\ref{appendix:two:renorm:mass}, we present a detailed discussion on the connection between
two renormalized quark masses introduced in LF quantization and equal-time quantization for the ``tetraquark" case.

\section{Setup of the stage}
\label{setup:stage}

The extended 't Hooft model contains both bosonic and fermionic quarks and gluons. For simplicity, we only
consider a single species of bosonic quark and fermionic quark.
The corresponding hybrid $\QCDtwo$ Lagrangian is dictated by the $\SUNc$ gauge invariance:
\begin{align}
\mathcal{L}_{h\QCDtwo} &= -\frac{1}{4}\left(F_{\mu\nu}^a\right)^2 + \bar{\psi} (i\!\!\not\! D-m_F) \psi
  + \left(D^{\mu}\phi\right)^{\dagger}D_{\mu}\phi - m_B^2\phi^{\dagger}\phi,
\label{Lag:extended:t:Hooft:model}
\end{align}
where $\psi$ and $\phi$ denote the Dirac and complex scalar fields, $m_F$ and $m_B$ refer to the masses of the bosonic and fermionic quarks,
and $A_{\mu}^a$ represents the gluon field with color index $a=1,2,\cdots, \Nc^2-1$.
$D_{\mu} = \partial_{\mu} - i g_{s} A_{\mu}^{a}T^a$ signifies the color covariant derivative,
and $F_{\mu\nu}^{a}\equiv\partial_{\mu}A_{\nu}^{a}-\partial_{\nu}A_{\mu}^{a}+g_{s}f^{abc}A_{\mu}^{b}A_{\nu}^{c}$ represents the gluonic field strength tensor.

The generators of $\SUNc$ group in the fundamental representation obey the following relation:
\bseq
\begin{align}
\mathrm{tr}(T^{a}T^{b}) &= \frac{\delta^{ab}}{2},
\\
\sum_{a}T_{ij}^{a}T_{kl}^{a} &= \frac{1}{2}\left(\delta_{il}\delta_{jk}-\frac{1}{N_c}\delta_{ij}\delta_{kl}\right).
\end{align}
\eseq

The Lorentz two-vector is defined as $x^{\mu}=\left(x^{0},x^{z}\right)$,
with the superscript $0$ and $z$ representing the time and spatial indices.
The Dirac $\gamma$-matrices in two space-time dimensions are represented by
\begin{align}
\gamma^{0}=\sigma_{1},\qquad
\gamma^{z}=-i\sigma_{2},\qquad
\gamma^{5}\equiv\gamma^{0}\gamma^{z}=\sigma_{3},
\end{align}
where $\sigma^i$ ($i=1,2,3$) signifies the Pauli matrices.

In LF quantization it is also convenient to adopt the light-cone coordinates,
which are defined through $x^\pm = x_\mp = \left(x^0 \pm x^z\right)/\sqrt{2}$,
with the light-front time denoted by $x^+$.

Throughout this work, we are interested in the large-$\Nc$ limit:
\begin{align}
  \Nc \rightarrow \infty,\qquad\qquad
\lambda \equiv \frac{\gs^2\Nc}{4\pi}\text{ fixed},
\end{align}
with $\lambda$ referring to the 't Hooft coupling constant.
 We are also tacitly working in the so-called weak coupling limit, where $m_F, m_B\gg g \sim 1/\sqrt{N_c}$ \cite{Zhitnitsky:1985um}.

\section{BSEs for ``tetraquark" from Hamiltonian approach
\label{sec:scalar}}

In this section we first revisit the derivation of BSE of a ``tetraquark" in IMF, then revisit the derivation of the BSEs of a ``tetraquark"  in FMF.
Some special attention is paid to the quark mass renormalization in both LF and equal-time quantization.

\subsection{BSE of ``tetraquark" in IMF}

We start with rederivation of the BSE of ``tetraquark" in the IMF using the operator approach~\cite{Tomaras:1978nt}.
For this purpose, it is most convenient to quantize the scalar $\QCDtwo$ in equal LF time.

\subsubsection{LF Hamiltonian of scalar $\QCDtwo$}

We express the scalar $\QCDtwo$ lagrangian in terms of light-cone coordinates. Similar to the original 't Hooft model,
a great simplification can be achieved by imposing the light-cone gauge $A^{+a}=0$~\footnote{For notational simplicity, we will use the symbol
$m$ instead of $m_B$ to signify the bosonic quark mass in this section.}:
\begin{align}
  \mathcal{L}_{\mathrm{sQCD}_2} &= \frac{1}{2}\left(\partial_{-}A^{-a}\right)^2
  + \left(\partial_-\phi^\dagger\right) D_+\phi + \left(D_+\phi\right)^\dagger \partial_-\phi
  - m^2\phi^\dagger\phi.
\label{eq:Lag:scalar:LC}
\end{align}
The canonical conjugate momenta of bosonic quark fields are given by $\pi \equiv \frac{\partial\mathcal{L}}{\partial\left(\partial_+\phi^\dagger\right)}
= \partial_-\phi$ and $\pi^\dagger = \partial_-\phi^\dagger$.
After Legendre transformation, one arrives at the following LF Hamiltonian:
\begin{align}
    H_\LF & =\int dx^-\left(-\frac{1}{2}\left(\partial_{-}A^{-a}\right)^{2}+ig_{s}A^{-a}\left(\pi^{\dagger}T^{a}\phi-\phi^{\dagger}T^{a}\pi\right)+m^{2}\phi^{\dagger}\phi\right).
\label{EQ:LF:Hambiltonian:first:sQCD2}
\end{align}

Due to the absence of the light-front time derivative of the gluon field in \eqref{eq:Lag:scalar:LC}, $A^{-a}$ is no longer a dynamical
variable, which is subject to the following constraint:
\beq
  \partial_-^2A^{-a} = \gs J^a,
\label{eq:A:EOM}
\eeq
with $J^{a}\equiv i\left(\phi^{\dagger}T^{a}\pi-\pi^{\dagger}T^{a}\phi\right)$.

Solving $A^{-a}$ in term of $J^a$ in \eqref{eq:A:EOM},  and substituting back into \eqref{EQ:LF:Hambiltonian:first:sQCD2},
one then reduces the LF Hamiltonian to
\begin{align}
  H_\LF & =\int dx^{-}\left(m^{2}\phi^{\dagger}\phi-\frac{g_{s}^{2}}{2}J^{a}\frac{1}{\partial_{-}^{2}}J^{a}\right).
\label{eq:hamiltonian}
\end{align}

The LF Hamiltonian now becomes nonlocal.
Note that the rigorous meaning of $1/\partial_-^2 J^a$ in
\eqref{eq:hamiltonian} is
\begin{align}
\frac{1}{\partial_-^2} J^a\left(x^-\right) &=
    \int dy^- G^{(2)}_\rho\left(x^--y^-\right) J^a\left(y^-\right),
\label{express:as:Green:func}
\end{align}
where $G^{(2)}$ represents the Green function
\begin{align}
\partial_{-}^{2} G^{\left(2\right)}\left(x^{-}\right)&=\delta\left(x^{-}\right).
\end{align}
The actual solution of the Green function turns out to be
\begin{align}
 G_\rho^{(2)}\left(x^--y^-\right) &
 = -\int_{-\infty}^{+\infty}
 \frac{dk^+}{2\pi}\Theta\left(\left|k^+\right|-\rho\right)
 \frac{e^{ik^+\left(x^--y^-\right)}}{\left(k^+\right)^2}.
\label{G2:actual:form}
\end{align}
To render $G^{(2)}$ mathematically well-defined,
we have introduced an infrared cutoff $\rho$ to regularize the severe IR divergence pertaining to $k^+\to 0$.
This parameter may also be viewed as an artificial gauge parameter.
Needless to say, this fictitious parameter must disappear in the final expressions for
any physical entities.

\subsubsection{Quantization and bosonization}
\label{LF:s:quant:boson}

We impose the canonical quantization for the scalar $\QCDtwo$ in \eqref{eq:hamiltonian} in equal LF time.
It is convenient to Fourier-expand the $\phi$ and $\pi$ fields in terms of the quark and antiquark's annihilation and creation operators:
\bseq
\begin{align}
\phi^i\left(x^{-}\right) &= \int_0^\infty\frac{dk^+}{2\pi}\frac{1}{\sqrt{2k^+}}
\left[a^{i}\left(k^{+}\right)e^{-ik^{+}x^{-}}+c^{i\dagger}\left(k^{+}\right)e^{ik^{+}x^{-}}\right],
\\
  \pi^{j\dagger}\left(x^{-}\right) &= i\int_0^\infty\frac{dk^+}{2\pi}
    \sqrt{\frac{k^+}{2}}\left[a^{j\dagger}\left(k^{+}\right)e^{ik^{+}x^{-}}-c^{j}\left(k^{+}\right)e^{-ik^{+}x^{-}}\right].
\end{align}
\label{eq:scalar:LC:quant}
\eseq
where $i,j=1,\cdots,\Nc$ are color indices. The annihilation and creation operators are assumed  to obey the
standard commutation relations:
\begin{align}
  \left[a^i\left(k^+\right), a^{j\dagger}\left(p^+\right)\right] =
  \left[c^i\left(k^+\right), c^{j\dagger}\left(p^+\right)\right] =
  \left(2\pi\right)\delta\left(k^+-p^+\right)\delta^{ij}.
\label{a:c:quantization}
\end{align}

A useful trick to diagonalize the Hamiltonian is the
bosonization technique~\cite{Kikkawa:1980dc,Nakamura:1981zi,Dhar:1994ib,Dhar:1994aw,Cavicchi:1993jh,Barbon:1994au,Itakura:1996bk}.
One first introduces the following four compound color-singlet operators:
\bseq
\begin{align}
W\left(k^{+},p^{+}\right) &\equiv\frac{1}{\sqrt{\Nc}}\sum_{i}c^{i}\left(k^{+}\right)a^{i}\left(p^{+}\right),
&
W^{\dagger}\left(k^{+},p^{+}\right) &\equiv\frac{1}{\sqrt{\Nc}}\sum_{i}a^{i\dagger}\left(p^{+}\right)c^{i\dagger}\left(k^{+}\right)
    \label{eq:scalar:bosonize:Ndg},
\\
A\left(k^{+},p^{+}\right) & \equiv\sum_{i}a^{i\dagger}\left(k^{+}\right)a^{i}\left(p^{+}\right),
&
C\left(k^{+},p^{+}\right) &\equiv\sum_{i}c^{i\dagger}\left(k^{+}\right)c^{i}\left(p^{+}\right).
\end{align}
\label{eq:scalar:bosonize}
\eseq
It is straightforward to find the commutation relations among these four compound operators:
\bseq
\begin{align}
  \left[W\left(k_{1}^{+},p_{1}^{+}\right),W^{\dagger}\left(k_{2}^{+},p_{2}^{+}\right)\right] &=
    \left(2\pi\right)^{2}\delta\left(k_{1}^{+}-k_{2}^{+}\right)\delta\left(p_{1}^{+}-p_{2}^{+}\right)+\mathcal{O}\left(\frac{1}{\Nc}\right),
\\
  \left[W\left(k_{1}^{+},p_{1}^{+}\right),A\left(k_{2}^{+},p_{2}^{+}\right)\right] &=
    2\pi\delta\left(p_{1}^{+}-k_2^+\right) W\left(k_1^+,p_{2}^{+}\right),
 \\
  \left[W\left(k_{1}^{+},p_{1}^{+}\right),C\left(k_{2}^{+},p_{2}^{+}\right)\right] &=
    2\pi\delta\left(k_1^+-k_2^+\right) W\left(p_{2}^{+},p_{1}^{+}\right),
    \\
\left[A\left(k_{1}^{+},p_{1}^{+}\right),A\left(k_{2}^{+},p_{2}^{+}\right)\right] &=
    2\pi\delta\left(p_{1}^{+}-k_2^+\right)A\left(k_1^+,p_{2}^{+}\right)
    - 2\pi\delta\left(p_{2}^{+}-k_1^+\right) A\left(k_2^+,p_{1}^{+}\right),
  \\
 \left[C\left(k_{1}^{+},p_{1}^{+}\right),C\left(k_{2}^{+},p_{2}^{+}\right)\right] &=
    2\pi\delta\left(p_{1}^{+}-k_2^+\right)C\left(k_1^+,p_{2}^{+}\right)
    - 2\pi\delta\left(p_{2}^{+}-k_1^+\right) C\left(k_2^+,p_{1}^{+}\right),
 \\
  \left[A\left(k_{1}^{+},p_{1}^{+}\right),C\left(k_{2}^{+},p_{2}^{+}\right)\right] &= 0.
\end{align}
\label{eq:boson:commute}
\eseq

Substituting \eqref{eq:scalar:LC:quant} into the LF Hamiltonian in \eqref{eq:hamiltonian}, and express everything in terms of the compound operator basis
as specified in \eqref{eq:scalar:bosonize}, we can break the light-front Hamiltonian into three pieces:
\begin{align}
H_{\rm LF} & =H_{\rm LF;0}+\colon H_{\rm LF;2}\colon+\colon H_{\rm LF;4}\colon+\mathcal{O}\left(\frac{1}{\sqrt{\Nc}}\right),
\label{H:LF:scalar:decomposition}
\end{align}
whose explicit expressions read
\bseq
\begin{align}
 & H_{\rm LF;0} = N_c \int \frac{dx^{-}}{2\pi}\Big(\int_0^\infty\frac{{m^{2}}{dk^{+}}}{2k^+}-{\pi\lambda}\int_0^\infty\frac{dk_{3}^{+}}{2\pi}\int_0^\infty
  \frac{dk_{4}^{+}}{2\pi}\frac{\left(k_{3}^{+}-k_{4}^{+}\right)^2}{\left(k_{3}^{+}+k_{4}^{+}\right)^{2}{k_3^{+}k_4^{+}}} \Theta\left(|k_{3}^{+}+k_{4}^{+}|-\rho\right)\Big),
\\
  & :H_{\rm LF;2}: = {m^{2}}\int_0^\infty\frac{dk^{+}}{2\pi{2k^+}}
    \left[A\left(k^{+},k^{+}\right)+C\left(k^{+},k^{+}\right)\right]
+\int_0^\infty\frac{dk_{1}^{+}}{2\pi}\int_{-\infty}^\infty\frac{dk_{2}^{+}}{2\pi}\frac{2\pi\lambda}{k_{1}^{+} |k_{2}^{+}|}\left(\frac{k_{1}^{+}+k_{2}^{+}}{k_{2}^{+}-k_{1}^{+}}\right)^{2}
\nn\\
&\qquad\times \Theta(|k_{2}^{+}-k_{1}^{+}|-\rho)[ A\left(k_{1}^{+},k_{1}^{+}\right)+ C\left(k_{1}^{+},k_{1}^{+}\right)]\label{eq:3},
\\
 &  :H_{\rm LF;4}: = -{\pi^{3}\lambda}
    \int_0^\infty\frac{dk_{1}^{+}}{2\pi\sqrt{k_{1}^{+}}}\int_0^\infty\frac{dk_{2}^{+}}{2\pi\sqrt{k_{2}^{+}}}
    \int_0^\infty\frac{dk_{3}^{+}}{2\pi\sqrt{k_{3}^{+}}}\int_0^\infty\frac{dk_{4}^{+}}{2\pi\sqrt{k_{4}^{+}}}
    \frac{1}{\left(k_{3}^{+}-k_{4}^{+}\right)^{2}}
\nn\\
    &\qquad\times\Theta\left(|k_{3}^{+}-k_{4}^{+}|-\rho\right)
    \left(k_{1}^{+}+k_{2}^{+}\right)\left(k_{4}^{+}+k_{3}^{+}\right)
\nn\\
    &\qquad\times\big[W^{\dagger}\left(k_{4}^{+},k_{1}^{+}\right)
    W\left(k_{3}^{+},k_{2}^{+}\right)
    \delta\left(k_{1}^{+}-k_{2}^{+}+k_{4}^{+}-k_{3}^{+}\right)
\nn\\
&\qquad + W^{\dagger}\left(k_{2}^{+},k_{3}^{+}\right)W\left(k_{1}^{+},k_{4}^{+}\right)
\delta\left(k_{2}^{+}-k_{1}^{+}+k_{3}^{+}-k_{4}^{+}\right)\big],
\label{eq:HLF4:sQCD2}
\end{align}
\label{eq:H89}
\eseq
with $:\;\::$ denoting the normal ordering. $H_{\rm LF;0}$ denotes the LF energy of the vacuum, which appears to be severely IR divergent.

The confinement characteristics of QCD indicates that all the physical excitation must be the color singlets.
In the color-singlet subspace of states, the compound operators $A$ and $C$ are not independent operators, which, at lowest order in $1/\Nc$,
actually can be expressed as the convolution between the color-singlet quark-antiquark pair
creation/annihilation operators $W$ and $W^{\dagger}$ in \eqref{eq:scalar:bosonize:Ndg}:
\bseq
\begin{align}
  A(k^+,p^+) &\rightarrow \int_0^\infty\frac{dq^+}{2\pi} W^\dagger(q^+,k^+)W(q^+,p^+),
\\
C(k^+,p^+) &\rightarrow \int_0^\infty\frac{dq^+}{2\pi} W^\dagger(k^+,q^+)W(p^+,q^+).
\end{align}
\label{eq:A:C:to:W:Wdagger}.
\eseq

Plugging \eqref{eq:A:C:to:W:Wdagger} into \eqref{eq:3} and \eqref{eq:HLF4:sQCD2}, and relabelling
the momenta in $W^{(\dagger)}\left(k^+,p^+\right)$ by $p^+=xP^+$, $k^+=(1-x)P^+$,
we can rewrite the :$H_{\rm LF;2}:$ and :$H_{\rm LF;4}$: pieces at the lowest order in $1/\Nc$ as
\bseq
\begin{align}
  :H_{\rm LF;2}: &= \int_{0}^{\infty} \frac{dP^{+}\,P^+}{(2\pi)^2}\int_0^1 dx
  W^{\dagger}\left(\left(1-x\right)P^{+},xP^{+}\right)W\left(\left(1-x\right)P^{+},xP^{+}\right)
\nn \\
  &\bigg[\frac{m^{2}}{2xP^{+}}+\frac{m^{2}}{2(1-x)P^{+}}
    +\frac{\lambda}{8xP^{+}}\int_{-\infty}^\infty\frac{dy}{|y|}
      \frac{\left(x+y\right)^2}{\left(y-x\right)^2}\Theta\left(|(y-x)P^{+}|-\rho\right)
\nn\\
  &+\frac{\lambda}{8(1-x)P^{+}}\int_{-\infty}^\infty\frac{dy}{|1-y|}
    \frac{\left(2-x-y\right)^2}{\left(y-x\right)^2}\Theta\left(|(y-x)P^{+}|-\rho\right)\bigg],
\\
  :H_{\rm LF;4}: &= -\frac{\lambda}{2\left(2\pi\right)^{2}}
    \int_0^1 dx\int_0^1 dy\int_{0}^{\infty} \frac{dP^+\left(P^{+}\right)^{2}}{2\pi} W^{\dagger}\left(\left(1-x\right)P^{+},xP^{+}\right)W\left(\left(1-y\right)P^{+},yP^{+}\right)
\nn\\
&\frac{1}{\sqrt{xP^{+}}}\frac{1}{\sqrt{yP^{+}}}\frac{1}{\sqrt{(1-y)P^{+}}}\frac{1}{\sqrt{(1-x)P^{+}}}\frac{1}{[(x-y)P^{+}]^{2}}
\nn \\
& \Theta\left(|(x-y)P^{+}|-\rho\right)\left[\left(1-x\right)P^{+}+\left(1-y\right)P^{+}\right]\left(yP^{+}+xP^{+}\right).
\end{align}
\label{eq:H:1}
\eseq

\subsubsection{Diagonalization of LF Hamiltonian}

Our strategy of deriving the BSE is by enforcing the light-front Hamiltonian \eqref{eq:H:1} in a diagonalize form.
For this purpose, we introduce an infinite tower of {\it tetraquark} annihilation/creation operators: $w_{n}(P^{+})$/$w_{n}^{\dagger}(P^{+})$, with $n$ and $P^{+}$
indicating the principal quantum number and the light-cone momentum of the ``meson" in the physical spectrum.
We assume that the $w_{n}(P^{+})$/$w_{n}^{\dagger}(P^{+})$ operator basis can be transformed into the color-singlet quark-antquark pair
creating/annihilation operator basis in the following fashion:
\bseq
\begin{align}
 W\left(\left(1-x\right)P^{+},xP^{+}\right) & =\sqrt{\frac{2\pi}{P^{+}}}\sum_{n=0}^{\infty}\chi_{n}\left(x\right)w_{n}\left(P^{+}\right),
\\
w_{n}\left(P^{+}\right) & =\sqrt{\frac{P^{+}}{2\pi}}\int_{0}^{1}dx\chi_{n}\left(x\right)W\left(\left(1-x\right)P^{+},xP^{+}\right),
\end{align}
\eseq
with the coefficient function $\chi_n(x)$ interpreted as the light-cone wave function of the $n$-th ``tetraquark".

It is desirable to demand that the ``tetraquark" annihilation and creation operators obey the standard commutation relations:
\begin{align}
   \left[w_{n}\left(P_1^{+}\right),w_{m}^{\dagger}\left(P_2^{+}\right)\right] & =2\pi\delta\left(P_1^{+}-P_2^{+}\right)\delta_{nm},
\end{align}
consequently the light-cone wave function $\chi_{n}(x)$ must satisfy the following orthogonality and completeness conditions:
\bseq
\begin{align}
\int_{0}^{1}dx\chi_{n}\left(x\right)\chi_{m}\left(x\right) & =\delta_{nm},
\\
\sum_{n}\chi_{n}\left(x\right)\chi_{n}\left(y\right) & =\delta\left(x-y\right).
\end{align}
\eseq

The $n$-th ``tetraquark" state can be constructed via
\beq
|P_n^{-},P^{+}\rangle  =\sqrt{2P^{+}}\, w_{n}^{\dagger}(P^{+})|0\rangle,
\eeq
where $P_n^{-}=M_{n}^{2}/(2P^{+})$ denotes the LF energy of the $n$-th excited ``tetraquark" state,
with $M_n$ the respective tetraquark mass.

In the $\Nc\to \infty$ limit, the scalar $\QCDtwo$ is composed of an infinite number of non-interacting
mesons. To account for this fact, one anticipates that the LF Hamiltonian
can be recast into a simple diagonal form in terms of the ``tetraquark" annihilation/creation operators:
\begin{align}
  H_{\rm LF} &= H_{\rm LF;0}   + \int_0^\infty\frac{dP^{+}}{2\pi}P_n^{-}\,
 \sum_n w_{n}^{\dagger}\left(P^{+}\right)w_{n}\left(P^{+}\right).
\label{eq:form}
\end{align}

In order to arrive at the desired form \eqref{eq:form}, all the non-diagonal terms in \eqref{eq:H:1} after transformed in the $w_{n}/w_{n}^{\dagger}$ basis, exemplified by
$w_n^\dagger w_m$ ($m\neq n$), $w^\dagger w^\dagger$, $w w$, $\cdots$,  must vanish.
This condition imposes some nontrivial constraint on the light-cone wave function $\chi_n(x)$,
which can be cast into an integral equation:
\begin{align}
  &\bigg(\frac{m^{2}}{x}+\frac{m^{2}}{1-x}+\frac{\lambda}{4x}\pint_{-\infty}^\infty\frac{dy}{|y|}\frac{\left(y+x\right)^{2}}{\left(y-x\right)^{2}}
+\frac{\lambda}{4(1-x)}\pint_{-\infty}^\infty\frac{dy}{|1-y|}\frac{\left(2-x-y\right)^{2}}{\left(y-x\right)^{2}}\bigg)\chi_{n}\left(x\right)
\nn \\
&-\frac{\lambda}{2}\pint_0^1\frac{dy}{\left(x-y\right)^{2}}\frac{\left(2-x-y\right)\left(x+y\right)}{\sqrt{x\left(1-x\right)y\left(1-y\right)}}
\chi_{n}\left(y\right)=M_{n}^{2}\chi_{n}\left(x\right).
\label{eq:scalar:LC:WE}
\end{align}
Reassuringly, the potential IR divergence as $y\rightarrow{x}$ is tamed by the principal value (PV) prescription, denoted by the symbol $-\mkern-16mu\int$. Note the occurrence
of the PV arises from taking the vanishing limit of the artificial IR regulator $\rho$ first introduced in \eqref{G2:actual:form}.
Here we show two PV prescriptions defined in term of the IR regulator $\rho$~\cite{Einhorn:1976uz,Mandelstam:1982cb,Leibbrandt:1987qv}:
\begin{align}
 \pint dy\frac{f\left(y\right)}{\left(x-y\right)^2}
 & \equiv \lim_{\rho\to 0^{+}}
    \int dy\frac{f\left(y\right)}{2}\left[\frac{1}{\left(x-y+i\rho\right)^2}
      +\frac{1}{\left(x-y-i\rho\right)^2}\right]
\nn \\
 & = \lim_{\rho\rightarrow0^{+}}\int dy\Theta\left(|x-y|-\rho\right)\frac{f\left(y\right)}{\left(x-y\right)^{2}}-\frac{2f\left(x\right)}{\rho}.
 \label{eq:PV}
\end{align}

\subsubsection{Quark mass renormalization and the renormalized BSE}

Though the IR divergence is cured by the PV prescription, the BSE in scalar
${\rm QCD}_2$, \eqref{eq:scalar:LC:WE}, is still plagued with logarithmic ultraviolet divergences, which arise
as $y\rightarrow{0}$ or $y\to \pm \infty$ in the first integral, and also arise as $y\rightarrow{1}$ or $y\to \pm \infty$ in the second integral in \eqref{eq:scalar:LC:WE}.

As first pointed out by Shei and Tsao~\cite{Tomaras:1978nt}, it is essential to renormalize the quark mass $m$ in order to eliminate the UV divergence.
Concretely speaking, one introduce the renormalized quark mass $m_r$ according to
\begin{align}
m_{r}^{2} & =m^{2}+\frac{\lambda}{2}\int_\delta^\Lambda\frac{dy}{y},
\label{mass:renormalization}
\end{align}
where the mass counterterm logarithmically depends on the UV cutoffs $\Lambda\gg \sqrt{\lambda}$,
and $\delta\to 0^+$.

Replacing the integration boundaries of the first two integrals on the left side of \eqref{eq:scalar:LC:WE}
with $\int_{-\Lambda}^{-\delta} + \int_{\delta}^{\Lambda}$ and $\int_{-\Lambda}^{1-\delta} + \int_{1+\delta}^{\Lambda}$, respectively, and working out the integrals,
it is straightforward to find that they diverge in the form of $\frac{\lambda}{2}\ln\Lambda/\delta$, which are
canceled exactly by the quark mass counterterm, leaving out a finite remnant $-2\lambda$.
Consequently the BSE becomes UV regular, which entails the renormalized quark mass only~\cite{Tomaras:1978nt}:
\begin{align}
  \left(\frac{m_{r}^{2}-2\lambda}{x}+\frac{m_{r}^{2}-2\lambda}{1-x}\right)\chi_{n}\left(x\right)
  -\frac{\lambda}{2}\pint_0^1\frac{dy}{\left(x-y\right)^{2}}
  \frac{\left(2-x-y\right)\left(x+y\right)}{\sqrt{x\left(1-x\right)y\left(1-y\right)}}
  \chi_{n}\left(y\right) = M_{n}^{2}\chi_{n}\left(x\right).
\label{eq:scalar:LC:WE:renorm}
\end{align}
Note that this BSE is similar to, albeit slightly more involved than, the celebrated 't Hooft's
equation in spinor $\QCDtwo$.

\subsection{BSE of ``tetraquark" in FMF}

It is advantageous to derive the BSE in $\QCDtwo$ in FMF in the familiar equal-time quantization.
The BSE in scalar $\QCDtwo$ was recently derived with the aid of the operator approach in equal-time quantization~\cite{Ji:2018waw}.
The goal of this subsection is essentially to revisit the derivation in \cite{Ji:2018waw}, with some new elements added.
For instance, we present a new way of deriving the mass gap equation from the variational perspective,
as well as elaborate on the subtlety pertaining to the quark mass renormalization.
Moreover, we also for the first time employ the diagrammatic technique to derive the BSE of a tetraquark in FMF.
We devote Appendix~\ref{Diagram:derivation:sQCD2} to a detailed explanation of deriving the BSE of ``tetraquark" in FMF from the perspective
of DS/BS equations.

\subsubsection{The Hamiltonian in the axial gauge}

Similar to the treatment of the spinor $\QCDtwo$~\cite{Bars:1977ud}, it is most convenient to choose the axial gauge $A^{a\,z}=0$ to quantize the scalar $\QCDtwo$ in equal time.
The Lagrangian then reduces to
\begin{align}
  \mathcal{L}_{\sQCDtwo} &=
    \frac{1}{2}(\partial_{z}A_0^a)^2
    + (D_0\phi)^{\dagger}D_0\phi
    - (\partial_z\phi^\dagger)\partial_z\phi
    - m^2\phi^{\dagger}\phi \label{eq:4}.
\end{align}

The conjugate momenta are $\pi=D_{0}\phi$, $\pi^{\dagger}=(D_{0}\phi)^{\dagger}$.
The Hamiltonian is obtained through the Legendre transformation:
\begin{align}
  H &= \int dz\left(|\partial_z\phi|^2 + m^2|\phi|^2 + \pi^\dagger\pi\right)
    + \frac{\gs^2}{2}\int dz\left(J^{a}\frac{-1}{\partial_{z}^{2}}J^{a}\right),
\label{eq:H2}
\end{align}
with $J^{a}=i\left(\phi^{\dagger}T^{a}\pi-\pi^{\dagger}T^{a}\phi\right)$.

Similar to \eqref{express:as:Green:func}, one can express ${1\over \partial_z^2} J^a$ as the convolution
between $J^a$ and the Green function $\widetilde{G}^{\left(2\right)}$, which is defined through
\begin{align}
\tilde{G}_{\rho}^{\left(2\right)}(z) & =-\int_{-\infty}^{+\infty}\frac{dk}{2\pi}\Theta\left(|k|-\rho\right)\frac{e^{ik z}}{k^2
}.
\end{align}
Analogous to \eqref{G2:actual:form}, we again introduce an IR regulator $\rho$ to make the Green function well-defined.

\subsubsection{Dressed quark basis, mass-gap equation, and quark mass renormalization}

One can conduct the equal-time quantization for the Hamiltonian as specified in \eqref{eq:H2}.
It is convenient to Fourier-expand the $\phi$ and $\pi^\dagger$ fields in the basis of the quark/antiquark's annihilation and creation operators:
\bseq
\begin{align}
\phi^i\left(z\right) & =\int\frac{dk}{2\pi\sqrt{2E_{k}}}e^{ikz}\left[a^{i}\left(k\right)+c^{i\dagger}\left(-k\right)\right],
\label{eq:b1}
\\
\pi^{j\dagger}\left(z\right) & =i\int\frac{dk}{2\pi}\sqrt{\frac{E_{k}}{2}}e^{-ikz}\left[a^{j\dagger}\left(k\right)-c^{j}\left(-k\right)\right],
\end{align}
\label{eq:scalar:ET:quant}
\eseq
where $E_k$ denotes the energy of a {\it dressed} quark, whose concrete dispersion relation will be determined by the mass-gap equation in the following.
The commutation relations between the quark annihilation and creation operators are the same as \eqref{a:c:quantization}, except
all the $+$-components are replaced with the $z$-components.

Analogous to the bosonizaion procedure adopted in the LF case in Sec.~\ref{LF:s:quant:boson},
we introduce the following four color-singlet compound operators:
\bseq
\begin{align}
W\left(p,q\right) & \equiv\frac{1}{\sqrt{\Nc}}\sum_{i}c^{i}\left(-p\right)a^{i}\left(q\right),&
W^{\dagger}\left(p,q\right)  &\equiv\frac{1}{\sqrt{\Nc}}\sum_{i}a^{i\dagger}\left(q\right)c^{i\dagger}\left(-p\right),
\\
A\left(p,q\right) & \equiv\sum_{i}a^{i\dagger}\left(p\right)a^{i}\left(q\right),&
C\left(p,q\right) &\equiv\sum_{i}c^{i\dagger}\left(-p\right)c^{i}\left(-q\right).
\end{align}
\label{eq:C}
\eseq
The commutation relations among these compound operators are identical to
\eqref{eq:boson:commute}, except all the $+$ component are replaced with the $z$-components.

Analogous to \eqref{H:LF:scalar:decomposition}, we decompose the Hamiltonian \eqref{eq:H2} into three pieces:
\begin{align}
H & =H_{0}+\colon H_{2}\colon+\colon H_{4}\colon+\mathcal{O}\left(\frac{1}{\sqrt{\Nc}}\right),
\label{sQCD:Ham}
\end{align}
whose explicit expressions read
\bseq
\begin{align}
 & H_{0}=  \Nc\int{dz}\Big(\int\frac{dk{(k^2+m^2)}}{2\pi\left(2E_k\right)}+\int\frac{dk{E_k}}{4\pi}
+\frac{\pi\lambda}{4}\int\frac{dk_1}{2\pi}\int\frac{dk_2}{2\pi}\frac{\left(E_{k_2}-E_{k_1}\right)^{2}}{\left(k_{1}-k_{2}\right)^{2}}\frac{1}{E_{k_1}E_{k_2}}\Big),
\label{eq:H000}
\\
& :H_{2}:  =\int\frac{dk}{2\pi}\widetilde{\Pi}^{+}\left(k\right)\left(A\left(k,k\right)+C\left(k,k\right)\right)
+\sqrt{N_{c}}\int\frac{dk}{2\pi}\widetilde{\Pi}^{-}\left(k\right)\left(W\left(k,k\right)+W^{\dagger}\left(k,k\right)\right),
\\
& :H_{4}: =\frac{\lambda}{32\pi^{2}}\iiiint dk_{1}dk_{2}dk_{3}dk_{4}\frac{\delta\left(k_{2}-k_{1}+k_{4}-k_{3}\right)}{\left(k_{4}-k_{3}\right)^{2}}\Theta\left(|k_4-k_3|-\rho\right)
\nn \\
&\qquad \times \Big[-2f_{+}\left(k_{1},k_{2}\right)f_{+}\left(k_{3},k_{4}\right)W^{\dagger}\left(k_{1},k_{4}\right)W\left(-k_{2},-k_{3}\right)
\nn \\
&\qquad +f_{-}\left(k_{1},k_{2}\right)f_{-}\left(k_{3},k_{4}\right)W^{\dagger}\left(k_{1},k_{4}\right)W^{\dagger}\left(k_{3},k_{2}\right)
\nn \\
&\qquad + f_{-}\left(k_{1},k_{2}\right)f_{-}\left(k_{3},k_{4}\right)W\left(k_{1},k_{4}\right)W\left(k_{3},k_{2}\right)\Big],
\end{align}
\label{eq:H3}
\eseq
with $\widetilde{\Pi}^{\pm}(k)$ and $f_{\pm}$ defined by
\bseq
\begin{align}
\widetilde{\Pi}^{\pm}\left(k\right) & \text{=}\frac{1}{2}\left(\frac{k^{2}+m^{2}}{E_{k}}\pm E_{k}\right)+\frac{\lambda}{4}\int dk_{1}\frac{\frac{E_{k_{1}}}{E_{k}}\pm\frac{E_{k}}{E_{k_1}}}{\left(k+k_{1}\right)^{2}}\Theta\left(|k+k_1|-\rho\right),
\\
f_{\pm}(k_{1},k_{2}) & =\sqrt{\frac{E_{k_2}}{E_{k_1}}}\pm\sqrt{\frac{E_{k_1}}{E_{k_2}}}.
\label{f:plus:minus}
\end{align}
\label{Pi:pm:f:pm:def}
\eseq

It is desirable to put the $:H_{2}:$ piece, which governs the dressed quark energy, into a diagonalized form.
For this purpose, the coefficient of the off-diagonal $W+W^{\dagger}$ term in $:H_{2}:$ is demanded to vanish.
The constraint $\widetilde{\Pi}^{-}=0$ then leads to a constraint for $E_k$~\cite{Ji:2018waw}:
\begin{align}
\frac{k^{2}+m^{2}}{E_{k}}-E_{k}+\frac{\lambda}{2}\pint dk_{1}\left(\frac{E_{k_{1}}}{E_{k}}-\frac{E_{k}}{E_{k_{1}}}\right)\frac{1}{\left(k+k_{1}\right)^{2}} & =0.
\label{eq:scalar:mass_gap}
\end{align}
This integral equation can be solved numerically to determine the dispersion relation.
Following the spinor ${\rm QCD}_2$ case~\cite{Bars:1977ud}, we also refer to this equation as
the mass-gap equation.

Here we provide an alternative route to derive the mass-gap equation \eqref{eq:scalar:mass_gap}.
The physical requirement is that the vacuum energy $H_0$ in \eqref{eq:H000}, albeit being severely divergent,
should be minimized with respect to all possible functional forms of $E_k$.
This requirement leads to a variational equation:
\begin{align}
\frac{\delta H_{0}[E_k]}{\delta E_{p}} & =0,
\end{align}
which leads to
\begin{align}
& -\int\frac{dk}{2\pi}\frac{k^2+m^2}{E_{k}^2}\delta(k-p)+\int\frac{dk}{2\pi}\delta(k-p)\nn\\
& +{\pi\lambda}\int\frac{dk_1}{2\pi}\int\frac{dk_2}{2\pi}\frac{1}{(k_1-k_2)^2}
\left(\frac{1}{E_{k_2}} - \frac{E_{k_2}}{E_{k_1}^2}\right)\delta(k_1-p) = 0.
\end{align}
Conducting the integration, we then recover the mass-gap equation in \eqref{eq:scalar:mass_gap}~\cite{Ji:2018waw}.

At first sight, the mass-gap equation \eqref{eq:scalar:mass_gap} suffers from both IR and UV divergences.
Actually, the potential IR divergence with $k_{1}\rightarrow-k$ can be tamed by the PV prescription
\eqref{eq:PV}. However, as $|k_{1}|\rightarrow\infty$, one can ascertain that $E_{k1}\rightarrow|k_1|$, and the integral
in \eqref{eq:scalar:mass_gap} exhibits a logarithmic UV divergence.
Fortunately, the UV divergence can be absorbed in the quark mass through the renormalization procedure.
One may follow \cite{Ji:2018waw} to introduce the renormalized quark mass:
\begin{align}
 m_{\tilde{r}}^{2} & =m^2+\frac{\lambda}{2}\pint dk_1\frac{E_{k_1}}{k_1^2}.
\label{quark:mass:counterterm:sQCD:ETQ}
\end{align}
Slightly differing from \cite{Ji:2018waw}, we have imposed the PV prescription to circumvent the IR divergence
arising from the $k_1\to 0$ region.

Plugging \eqref{quark:mass:counterterm:sQCD:ETQ} into \eqref{eq:scalar:mass_gap}, we then obtain the
renormalized mass-gap equation:
\begin{align}
  \frac{k^{2}+m_{\tilde{r}}^{2}}{E_{k}}-E_{k}
  + \frac{\lambda}{2}\pint dk_1 \left[
    \left(\frac{E_{k_{1}}}{E_{k}}-\frac{E_{k}}{E_{k_{1}}}\right)
    \frac{1}{\left(k+k_{1}\right)^{2}}
    -\frac{E_{k_1}}{E_{k}}\frac{1}{k_{1}^2}
    \right] &= 0,
\label{eq:scalar:mass_gap:renorm}
\end{align}
which is free from logarithmic UV divergence.

Since the prescribed renormalization scheme in equal-time quantization differs from that in the LF quantization,
the renormalized mass $m_{\tilde{r}}^{2}$ is not necessarily equal to $m_{r}$ introduced in \eqref{mass:renormalization}.
We devote Appendix~\ref{appendix:two:renorm:mass} to a detailed discussion on the connection between
these two renormalized quark masses.

\subsubsection{Bogoliubov transformation and diagonalization of Hamiltonian}

Following the same line of reasoning that leads to \eqref{eq:A:C:to:W:Wdagger}, the confinement feature of QCD implies that, at the lowest order
in $1/\Nc$,  the compound operators $A$ and $C$ defined in \eqref{eq:C} can be expressed as
\begin{align}
A\left(k_{1},k_{2}\right) & \rightarrow\intop\frac{dp}{2\pi}W^{\dagger}\left(p,k_{1}\right)W\left(p,k_{2}\right),
\qquad
C\left(k_{1},k_{2}\right)  \rightarrow\intop\frac{dp}{2\pi}W^{\dagger}\left(k_{1},p\right)W\left(k_{2},p\right).
\end{align}

With this replacement, the Hamiltonian \eqref{sQCD:Ham} can be built solely out of the  color-singlet compound operators $W$ and $W^\dagger$:
\begin{align}
H & =\int\frac{dpdq}{4\pi^{2}}\left(\widetilde{\Pi}^{+}\left(p\right)+\widetilde{\Pi}^{+}\left(q\right)\right)W^{\dagger}\left(p,q\right)W\left(p,q\right)
\nn \\
 & -\frac{\lambda}{32\pi^{2}}\int dP\iint dkdp\frac{O+Q}{(p-k)^{2}}\Theta\left(|p-k|-\rho\right),
\label{eq:HHH}
\end{align}
where
\bseq
\begin{align}
&O=2S_{+}\left(p,k,P\right)W^{\dagger}(P-p,p)W(P-k,k),
\\
&Q=S_{-}\left(p,k,P\right)\left(W^{\dagger}\left(p,P-p\right)W^{\dagger}\left(P-k,k\right)+W\left(p,P-p\right)W\left(P-k,k\right)\right),
\end{align}\eseq
with
\beq
S_{\pm}\left(p,k,P\right)=f_{\pm}\left(P-p,P-k\right)f_{\pm}\left(p,k\right).
\eeq

To put the Hamiltonian \eqref{eq:HHH} in a diagonalized form, it is advantageous to employ the
Bogoliubov transformation~\cite{Kalashnikova:2001df}. The key is to introduce a new set
of color-singlet ``tetraquark" operators represented by $w$ and $w^\dagger$. Schematically, two sets of color-singlet
operators are connected through the following Bogoliubov transformation:
\begin{align}
w=\mu W+ \nu W^{\dagger},
\nn \\
w^{\dagger}=\mu W^{\dagger} + \nu W, \nn \\
  \mu^2 -\nu^2 =1.
\label{eq:1.11}
\end{align}
The coefficient $\mu$ and $\nu$ can be determined such that the Hamiltonian gets diagonalized in the new operator basis.

In the case of scalar $\QCDtwo$, we introduce two infinite towers of color-singlet ``tetraquark" annihilation and creation operators,
$w_{n}$ and $w_{n}^{\dagger}$, which are linear combinations of the $W$ and $W^{\dagger}$ operators through Bogoliubov transformation.
Inversely, we can express the $W$ operators in terms of infinite sum of $w_{n}$ and $w_n^\dagger$ operators:
\begin{align}
W\left(q-P,q\right) & =\sqrt{\frac{2\pi}{|P|}}\sum_{n=0}^{\infty}\left[w_{n}\left(P\right)\chi_{n}^{+}\left(q,P\right)-w_{n}^{\dagger}
\left(-P\right)\chi_{n}^{-}\left(q-P,-P\right)\right].
\label{eq:hhh}
\end{align}

The operators $w_{n}(P)$ and $w_{n}^{\dagger}(P)$ bear clear physical meaning, which represent the annihilation and creation
operators for the $n$th ``tetraquark" state carrying momentum $P$.
The coefficient functions $\chi_{n}^{\pm}$ can be interpreted as the forward/backward-moving ``tetraquark" wave functions,
playing the role of the Bogoliubov coefficients $\mu$ and $\nu$ in \eqref{eq:1.11}.
The operators $w_{n}$ and $w_{n}^{\dagger}$ are anticipated to obey the standard commutation relations:
\bseq
\begin{align}
\left[w_{n}(P), w_{m}^{\dagger}(P^{'})\right] & =2\pi\delta_{nm}\delta\left(P-P^{'}\right),
\\
\left[w_{n}(P),w_{m}(P^{'})\right] & =\left[w_{n}^{\dagger}(P),w_{m}^{\dagger}(P^{'})\right]=0.
\end{align}
\eseq
To fulfill these commutation relations, the ``tetraquark" wave functions $\chi_n^{\pm}$ must satisfy the
following orthogonality and completeness conditions
\bseq
\begin{align}
\int_{-\infty}^{+\infty}dp\left[\chi_{+}^{n}\left(p,P\right)\chi_{+}^{m}\left(p,P\right)-\chi_{-}^{n}\left(p,P\right)\chi_{-}^{m}\left(p,P\right)\right] & =|P|\delta^{nm},
\\
\int_{-\infty}^{+\infty}dp\left[\chi_{+}^{n}\left(p,P\right)\chi_{-}^{m}\left(p-P,-P\right)-\chi_{-}^{n}\left(p,P\right)\chi_{+}^{m}\left(p,P\right)\right] & =0,
\\
\sum_{n=0}^{\infty}\left[\chi_{+}^{n}(p,P)\chi_{+}^{n}\left(q,P\right)-\chi_{-}^{n}\left(p-P,-P\right)\chi_{-}^{n}\left(q-P,-P\right)\right] & =|P|\delta\left(p-q\right),
\\
\sum_{n=0}^{\infty}\left[\chi_{+}^{n}\left(p,P\right)\chi_{-}^{n}\left(q,P\right)-\chi_{-}^{n}\left(p-P,-P\right)\chi_{+}^{n}\left(q-P,-P\right)\right] & =0.
\end{align}
\eseq

The physical vacuum is defined by $w_n(P)|\Omega\rangle=0$, Note $\vert \Omega\rangle$ differs from the dressed quark vacuum $\vert 0\rangle$,
which is defined by minimizing $:H_2:$.
The $n$-th mesonic state can be constructed by
\begin{align}
|P_n^{0},P\rangle & =\sqrt{2P_n^{0}}w_{n}^{\dagger}\left(P\right)|\Omega\rangle,
\end{align}
where $P_n^{0}=\sqrt{M_n^2+P^2}$ denotes the energy of the $n$-th mesonic state.

Plugging \eqref{eq:hhh} into \eqref{eq:HHH}, we desire to put the Hamiltonian in a diagonalized form
in terms of the ``tetraquark" creation and annihilation operators:
\begin{align}
H=H_0^{'}+\int\frac{dP}{2\pi}\sum_n P_n^{0}w_n^{\dagger}(P)w_n(P)+ \mathcal{O}\left(\frac{1}{\sqrt{\Nc}}\right),
\end{align}
where $H_0^{\prime}$ is the shifted vacuum energy.

Similar to the recipe leading to \eqref{eq:form} in the light-front case, we enforce that all the non-diagonal operators
\eqref{eq:HHH} in the new $w_{n}/w_{n}^{\dagger}$ basis, such as
$w_n^\dagger w_m$ ($m\neq n$), $w^\dagger w^\dagger$, $w w$, $\cdots$,  ought to vanish.
This criterion imposes quite nontrivial constraints on the mesonic wave function $\chi^\pm_n(x)$.
As a matter of fact, such constraint can be cast into two coupled integral equations for
$\chi_n^\pm$:
\begin{align}
& \left(\Pi^{+}\left(p\right)+\Pi^{+}\left(P-p\right)\mp P_n^{0}\right)\chi_{n}^{\pm}\left(p,P\right) \nn \\
& =\frac{\lambda}{4}\pint\frac{dk}{\left(p-k\right)^{2}}\times\left(S_{+}\left(p,k,P\right)\chi_{n}^{\pm}\left(k,P\right)-S_{-}\left(p,k,P\right)\chi_{n}^{\mp}\left(k,P\right)\right),
\label{eq:scalar:ET:WE}
\end{align}
where 
\begin{align}
\Pi^{+}\left(k\right) &=\widetilde{\Pi}^{+}\left(k\right) - \frac{\lambda}{\rho} =\frac{1}{2}\left(\frac{k^{2}+m^{2}}{E_{k}}+ E_{k}\right)+\frac{\lambda}{4}\pint dk_{1}\frac{\frac{E_{k_{1}}}{E_{k}}+\frac{E_{k}}{E_{k_1}}}{\left(k+k_{1}\right)^{2}},
\end{align}
which agree with the ``tetraquark" BSEs in scalar ${\rm QCD}_2$ in FMF recently derived by Ji, Liu and Zahed~\cite{Ji:2018waw}.

Some remarks are in order. As proved in the spinor $\QCDtwo$ case by Bars and Green~\cite{Bars:1977ud},
when boosted to IMF, {\it i.e.}, taking $P \to \infty$ limit, the backward-moving mesonic wave function
dies out, whereas the forward-moving mesonic wave function approaches the light-cone
wave function , consequently Bars-Green equations will reduce to 't Hooft equation.
This pattern perfectly fits into the tenet of LaMET, and one naturally anticipates the same
story will repeat itself for scalar $\QCDtwo$. Indeed, as formally proved in \cite{Ji:2018waw},
when boosted to IMF, the backward-moving ``tetraquark" wave function $\chi^-_n$ does fade away, while the forward-moving ``tetraquark" wave function $\chi^+_n$ approaches the light-cone wave function
$\chi_n(x)$ with $x \equiv p/P$. As a consequence, in the IMF the equal-time BSEs \eqref{eq:scalar:ET:WE} descend
to the Shei-Tsao equation in \eqref{eq:scalar:LC:WE:renorm}. In Sec.~\ref{sec:numerical} we will provide numerical evidence for the
aforementioned pattern, {\it viz.}, the forward-moving ``tetraquark" wave function indeed tends
to converge to its light-cone counterpart with the increasing ``tetraquark" momentum.

\section{BSEs for ``baryon" from Hamiltonian approach}
\label{sec:hybrid}

As mentioned in Introduction, if the bosonic quark can be interpreted as the diquark, the
bound state formed by the fermionic quark and bosonic antiquark in the hybrid $\QCDtwo$ may
bear some resemblance with the ordinary baryon in the real world. The goal of this section
is to derive the BSEs of such a ``baryon" in both IMF and FMF.

\subsection{BSE of ``baryon" in IMF}

The BSE of ``baryon" in the hybrid QCD2 was first obtained using diagrammatic approach
in LF quantization by Aoki in 1993~\cite{Aoki:1993ma}. Shortly after the ``baryon" mass spectra were also
studied by Aoki and Ichihara~\cite{Aoki:1995dh}. Note this BSE of ``baryon" is valid only in the IMF. In
this subsection, we will revise of the derivation of the ``baryon" BSE in IMF~\cite{Aoki:1993ma}, yet instead
starting from the Hamiltonian approach.

\subsubsection{LF Hamiltonian of hybrid  $\QCDtwo$}

We start from the hybrid $\QCDtwo$ Lagrangian which contains both scalar and spinor quarks.
Imposing the light-cone gauge $A^{+a}=0$ and adopting the light-front coordinates,
equation~\eqref{Lag:extended:t:Hooft:model} reduces to
\begin{align}
  \mathcal{L}_{\hQCDtwo} &=
    \frac{1}{2}\left(\partial_{-}A^{-a}\right)^{2}
    + i\left(\psi_{R}^{\dagger}D_{+}\psi_{R}+\psi_{L}^{\dagger}\partial_{-}\psi_{L}\right)
    - \frac{m_F}{\sqrt{2}}\left(\psi_{L}^{\dagger}\psi_{R}+\psi_{R}^{\dagger}\psi_{L}\right)\nn \\
    &\qquad+ \left(\partial_-\phi^\dagger\right) D_+\phi + \left(D_+\phi\right)^\dagger \partial_-\phi - m_B^{2}\phi^{\dagger}\phi.
\end{align}
with $m_B$ and $m_F$ signifying the masses of the bosonic and fermionic quarks, respectively.
Two sets of canonical momenta are $\pi_\phi=\partial_-\phi, \pi_\psi=\psi_R$. Note $\psi_L$ is a constrained rather than canonical variable.
After Legendre transformation, we obtain the following LF Hamiltonian:
\begin{align}
  H_\text{LF} & =
   \int dx^- \Big[ \frac{m_F}{\sqrt{2}}\psi_{R}^{\dagger}\psi_{L}+ m_B^{2}\phi^{\dagger}\phi
    - \frac{\gs^2}{2}J^a\frac{1}{\partial_{-}^{2}}J^a\Big],
\label{LF:Hamil:hybrid}
\end{align}
with
\beq
  J^a \equiv i\left(\phi^{\dagger}T^{a}\pi-\pi^{\dagger}T^{a}\phi\right)
    + \psi_{R}^{\dagger}T^{a}\psi_{R}.
\eeq

\subsubsection{Compound operators and fermionization}
\label{Compound:Fermionization}

The bosonic quark field has been Fourier-expanded in the quark annihilation/creation
operator basis in \eqref{eq:scalar:LC:quant}.
The fermionic quark field can be Fourier-expanded accordingly,
\begin{align}
\psi_{R}^{i}(x^{-}) & =\int_0^\infty\frac{dk^{+}}{2\pi}[b^{i}\left(k^{+}\right)e^{-ik^{+}x^{-}}+d^{i^{\dagger}}\left(k^{+}\right)e^{ik^{+}x^{-}}].
\end{align}

Following the bosonization procedure in the preceding section, here we introduce a set of color-singlet compound operators composed of the
bosonic and fermionic quark annihilation and creation operators.
Since the system we are studying is the ``baryon",
we refer to this procedure as {\it fermionization}.

Besides the bosonic compound operators already introduced in \eqref{eq:scalar:bosonize}, we enumerate some new color-singlet compound operators~\footnote{Note here
the normalization convention for the compound operators $B$ and $D$ follows that of \cite{Jia:2024atq}, where they are scaled by a factor of $\sqrt{\Nc}$ compared to those defined in \cite{Jia:2018qee}.}:
\bseq
\begin{align}
B\left(k^{+},p^{+}\right) & \equiv\sum_{i}b^{i\dagger}\left(k^{+}\right)b^{i}\left(p^{+}\right),
&
D\left(k^{+},p^{+}\right)  &\equiv\sum_{i}d^{i\dagger}\left(k^{+}\right)d^{i}\left(p^{+}\right), \\
K\left(k^{+},p^{+}\right) &\equiv\frac{1}{\sqrt{\Nc}}\sum_{i}b^{i}\left(p^{+}\right)c^{i}\left(k^{+}\right),
&
\overline{K}\left(k^{+},p^{+}\right) & \equiv\frac{1}{\sqrt{\Nc}}\sum_{i}d^{i}\left(k^{+}\right)a^{i}\left(p^{+}\right).
\end{align}
\label{eq:hsf}
 \eseq

Note the compound operator $K$ annihilates a fermionic quark and a  bosonic antiquark,
whereas the compound operator $\overline{K}$ annihilates a  fermionic antiquark and a bosonic quark.
The anti-commutation relations among the fermionic compound operators $K$, $K^{\dagger}$, $\overline{K}$ and $\overline{K}^{\dagger}$ become
\bseq
\begin{align}
  \left\{K\left(k_{1}^{+},p_{1}^{+}\right),
  K^{\dagger}\left(k_{2}^{+},p_{2}^{+}\right)\right\} &=
    \left(2\pi\right)^{2}\delta\left(k_{1}^{+}-k_{2}^{+}\right)
      \delta\left(p_{1}^{+}-p_{2}^{+}\right)
    + \mathcal{O}(1/N_c), \\
  \left\{\overline{K}\left(k_{1}^{+},p_{1}^{+}\right),
  \overline{K}^{\dagger}\left(k_{2}^{+},p_{2}^{+}\right)\right\}  & =
    \left(2\pi\right)^{2}\delta\left(k_{1}^{+}-k_{2}^{+}\right)
      \delta\left(p_{1}^{+}-p_{2}^{+}\right)
    + \mathcal{O}(1/N_c), \\
  \left\{K\left(k_{1}^{+},p_{1}^{+}\right),
   \overline{K}\left(k_{2}^{+},p_{2}^{+}\right)\right\} &=
    \left\{K^{\dagger}\left(k_{1}^{+},p_{1}^{+}\right),
   \overline{K}\left(k_{2}^{+},p_{2}^{+}\right)\right\} = 0, \\
  \left\{K\left(k_{1}^{+},p_{1}^{+}\right),\overline{K}^{\dagger}
  \left(k_{2}^{+},p_{2}^{+}\right)\right\} &=
    \left\{K^{\dagger}\left(k_{1}^{+},p_{1}^{+}\right),
   \overline{K}^{\dagger}\left(k_{2}^{+},p_{2}^{+}\right)\right\} = 0.
\end{align}
\eseq

We are particularly interested in the interaction term in \eqref{LF:Hamil:hybrid} that couples
the bosonic quark sector with the fermionic sector, which can be expressed in terms of the product of the
fermionic compound operators:
\begin{alignat}{1}
& -\left[i\gs\left(\phi^{\dagger}T^{a}\pi  \nn
-\pi^{\dagger}T^{a}\phi\right)\right]\frac{1}{\partial_{-}^{2}}  \nn \left(\gs\psi_{R}^{\dagger}T^{a}\psi_{R}\right)    \nn \\
& =-4\pi\lambda\int_{0}^{\infty}\frac{dk_{1}^{+}}{2\pi\sqrt{2k_{1}^{+}}}  \nn
\int_{0}^{\infty}\frac{dk_{2}^{+}}{2\pi\sqrt{2k_{2}^{+}}}\int_{0}^{\infty}\frac{dk_{3}^{+}}{2\pi} \nn \int_{0}^{\infty}\frac{dk_{4}^{+}}{2\pi}\frac{(k_{1}^{+}+k_{2}^{+})}{\left(k_{3}-k_{4}\right)^{2}}\Theta\left(|k_3-k_4|-\rho\right)  \nn\\
& \times 2\pi\delta\left(k_{2}^{+}-k_{1}^{+}+k_{3}^{+}-k_{4}^{+}\right)\left[ \overline{K}^{\dagger}\left(k_{4}^{+},k_{1}^{+}\right)\overline{K}\left(k_{3}^{+},k_{2}^{+}\right)+K^{\dagger}\left(k_{2}^{+},k_{3}^{+}\right)K\left(k_{1}^{+},k_{4}^{+}\right)
\right].
\end{alignat}

We then break the full LF Hamiltonian \eqref{LF:Hamil:hybrid} into three pieces:
\beq
H_{\rm LF} =H_{\rm LF;0}+\colon H_{\rm LF;2}\colon+\colon H_{\rm LF;4}\colon+
\mathcal{O}(1/\Nc),
\label{Breaking:LF:Hamiltonian:baryon}
\eeq
with
\bseq
\begin{align}
H_{\text{LF};0}  &= N_c\int\frac{dx^{-}}{2\pi}\Bigg(\int_{0}^{\infty}\frac{{m_{B}^{2}}dk^{+}}{2k^+}+\pi\lambda\int_{0}^{\infty}
\frac{dk_{3}^{+}}{2\pi}\int_{0}^{\infty}\frac{dk_{4}^{+}}{2\pi}\frac{\left(k_{3}^{+}-k_{4}^{+}\right)\left(k_{4}^{+}-k_{3}^{+}\right)}
{\left(k_{3}^{+}+k_{4}^{+}\right)^{2}{k_3^{+}k_4^{+}}}  \nn\\
&\times\Theta\left(|k_{3}^{+}+k_{4}^{+}|-\rho\right)+\frac{\lambda}{2}+\frac{\lambda-m_{F}^2}{2}\int_{\rho}^{\infty}\frac{dk^{+}}{k^{+}}\Bigg), \\
:H_{\text{LF};2}: &= \int\frac{dx^{-}}{2\pi}\int_{0}^{\infty}\frac{dk^{+}}{2k^{+}}\left[
m_B^2 A\left(k^{+},k^{+}\right)+m_B^2 C\left(k^{+},k^{+}\right)
+m_F^2 B\left(k^{+},k^{+}\right)+m_F^2 D\left(k^{+},k^{+}\right)
\right] \nn\\
  &+ \lambda\int_{0}^{\infty}\frac{dk^{+}}{2\pi}\left(\frac{1}{\rho}-\frac{1}{k^{+}}\right)\left[B\left(k^{+},k^{+}\right)+D\left(k^{+},k^{+}\right)\right]\nn \\
  &+ \frac{\lambda}{4}\int_{-\infty}^{\infty} dx^{-}\iint\frac{dk_{1}^{+}\cdot dk_{2}^{+}}
    {\left(2\pi\right) k_{1}^{+}k_{2}^{+}}
    \left(\frac{k_{1}^{+}+k_{2}^{+}}{k_{2}^{+}-k_{1}^{+}}\right)^{2}\Theta\left(|k_2-k_1|-\rho\right)\left[A\left(k_{1}^{+},k_{1}^{+}\right)+C\left(k_{2}^{+},k_{2}^{+}\right)\right], \label{eq:a1} \\
  :H_{\text{LF};4}: &= -\gs^2\Nc\int_{0}^{\infty}\frac{dk_{1}^{+}}{2\pi\sqrt{2k_{1}^{+}}}
    \int_{0}^{\infty}\frac{dk_{2}^{+}}{2\pi\sqrt{2k_{2}^{+}}}
    \int_{0}^{\infty}\frac{dk_{3}^{+}}{2\pi}\int_{0}^{\infty}\frac{dk_{4}^{+}}{2\pi}
    \frac{\left(k_{1}^{+}+k_{2}^{+}\right)}{\left(k_{3}-k_{4}\right)^{2}}\Theta\left(|k_3-k_4|-\rho\right)\nn \\
    &\qquad \times2\pi\delta\left(k_{2}^{+}-k_{1}^{+}+k_{3}^{+}-k_{4}^{+}\right)\left[\overline{K}^{\dagger}\left(k_{4}^{+},k_{1}^{+}\right)\overline{K}\left(k_{3}^{+},k_{2}^{+}\right)
    +K^{\dagger}\left(k_{2}^{+},k_{3}^{+}\right)K\left(k_{1}^{+},k_{4}^{+}\right)\right].
\label{eq:a2}
\end{align}
\label{eq:H0}
\eseq

As dictated by the confinement property of QCD, the bosonic color-singlet operators
$A$, $D$, $B$ and $C$ are not independent,
but can be replaced by the convolution of the following fermionic compound operators:
\bseq
\begin{align}
  A\left(p^+,q^+\right) &\rightarrow
    \int_{0}^{\infty}\frac{dr^+}{2\pi}
    \overline{K}^+\left(r^+,p^+\right)\overline{K}\left(r^+,q^+\right),
\\
D\left(p^+,q^+\right) &\rightarrow
    \int_{0}^{\infty}\frac{dr^+}{2\pi}
    \overline{K}^+\left(p^+,r^+\right)\overline{K}\left(q^+,r^+\right),
\\
  C\left(p^+,q^+\right) &\rightarrow
    \int_{0}^{\infty}\frac{dr^+}{2\pi}
    K^+\left(p^+,r^+\right)K\left(q^+,r^+\right),
\\
B\left(p^+,q^+\right) &\rightarrow
    \int_{0}^{\infty}\frac{dr^+}{2\pi}
    K^+\left(r^+,p^+\right)K\left(r^+,q^+\right).
\end{align}
\eseq

Substituting these relations to \eqref{eq:a1} and \eqref{eq:a2},  relabelling the fermionic antiquark and a bosonic quark of the light-cone momenta by
$p^{+}=xP^{+}$and $r^{+}=(1-x)P^{+}$,  and only retaining the leading-order terms in 1/$\Nc$, the LF Hamiltonian
can be solely built out of the fermionic compound operators $K$, $K^\dagger$, $\overline{K}$ and $\overline{K}^\dagger$:
\bseq
\begin{align}
:H_{\rm LF;2}: & =\int_{0}^{\infty} \frac{dP^{+}}{(2\pi)^{2}}\int_{0}^{1} dx K^{\dagger}((1-x)P^{+},xP^{+})K((1-x)P^{+},xP^{+})
\nn \\
&\times \left[ \frac{m_B^{2}}{2(1-x)}+\frac{m_F^{2}}{2x}+\frac{P^{+}\lambda}{\rho}-\frac{\lambda}{x}+\frac{1}{8(1-x)}\int_{-\infty}^\infty\frac{dy}{ |y|}\frac{(2-x-y)^2}{(y-x)^2}\Theta\left(|y-x|-\frac{\rho}{P^+}\right)\right]
\nn\\
& + (K\to \overline{K}),
\\
:H_{\rm LF;4}: & =-\frac{\lambda}{8\pi^{2}}\int_{0}^{\infty} dP^{+}\iint_{0}^{1} dxdy\overline{K}^{\dagger}\left((1-x)P^{+},xP^{+}\right)\overline{K}\left((1-y)P^{+},yP^{+}\right)
\nn \\
&\times \frac{1}{\sqrt{(1-x)(1-y)}}\frac{2-x-y}{(y-x)^{2}}\Theta\left(|y-x|-\frac{\rho}{P^+}\right)+ (K\to \overline{K}),
\end{align}
\label{LF:Ham:Hybrid:intermediate}
\eseq
where the charge conjugation symmetry has been invoked to condense the expression.

\subsubsection{Diagonalization of LF Hamiltonian}

Our goal is to diagonalize the LF Hamiltonian in \eqref{Breaking:LF:Hamiltonian:baryon}.
For this purpose, it is advantageous to introduce a new set of operators $k_{n}(P^{+})$/$k_{n}^{\dagger}(P^{+})$ (together with
$\bar{k}_{n}(P^{+})$/$\bar{k}_{n}^{\dagger}(P^{+})$), which
annihilates/creates the $n$-th ``baryon" (``anti-baryon") state, with $n$ indicating the
principal quantum number, and $P^{+}$ denotes
the light-cone momentum of the corresponding ``baryon" (``anti-baryon").
We hypothesize that the $K$ basis and the $k_{n}$ basis are connected through the following relation~\footnote{Obviously the charge conjugation invariance
indicates that $\overline{K}$ basis and the $\bar{k}_{n}$ basis are connected through the same relation, but with the coefficient functions
interpreted as the light-cone wave functions of the ``anti-baryon".
Since the ``baryon" and ``anti-baryon" degrees of freedom decouple in the
LF quantization, as is evident in \eqref{LF:Ham:Hybrid:intermediate}, we will concentrate on the ``baryon" BSE in this subsection.
The ``anti-baryon" BSE can be straightforwardly obtained via charge conjugation invariance.}:
\begin{align}
K\left(\left(1-x\right)P^{+},xP^{+}\right) & =\sqrt{\frac{2\pi}{P^{+}}}\sum_{n=0}^{\infty}\varPhi_{n}\left(x\right)k_{n}\left(P^{+}\right),
\end{align}
where the coefficient function $\varPhi_{n}(x)$ stands for the light-cone wave function of the $n$-th excited ``baryon" state.

If we demand that the ``baryon" annihilation/creation operators $k_n/k_n^\dagger$ obey
the standard anti-commutation relation:
\begin{align}
\left\{k_{n}\left(P_{1}^{+}\right),k_{m}^{\dagger}\left(P_{2}^{\text{+}}\right)\right\}  & =2\pi\delta\left(P_1^{+}-P_2^{+}\right)\delta_{nm},
\end{align}
the light-cone wave functions $\varPhi_{n}(x)$ must satisfy the following orthogonality and
completeness conditions:
\bseq
\begin{align}
& \int_{0}^{1}dx\,\varPhi_{n}\left(x\right)\varPhi^*_{m}\left(x\right)  =\delta_{nm},
\\
& \sum_{n}\varPhi_{n}\left(x\right)\varPhi^*_{n}\left(y\right) =\delta\left(x-y\right).
\end{align}
\eseq

The $n$-th ``baryon" state can be constructed via
\begin{align}
|P_n^{-},P^{+}\rangle & =\sqrt{2P^{+}}k_{n}^{\dagger}(P^{+})|0\rangle,
\end{align}
with the light-cone energy $P^{-}_n=M_{n}^{2}/(2P^{+})$, and $M_{n}$ represents
the mass of the $n$-th excited ``baryon" state.

In order to ultimately put the LF Hamiltonian in the desired diagonalized form:
\begin{align}
H_{\rm LF} & =H_{\rm LF;0}+\int\frac{dP^{+}}{2\pi}P_n^{-} \,\sum_n \Big[ k_{n}^{\dagger}(P^{+})k_{n}(P^{+})+
\bar{k}_{n}^{\dagger}(P^{+}) \bar{k}_{n}(P^+) \Big].
\end{align}
all the non-diagonal terms in \eqref{LF:Ham:Hybrid:intermediate}, after transformed into the
$k_{n}/k_{n}^{\dagger}$ basis, such as $k_n^\dagger k_m$ ($m\neq n$),
$k^\dagger k^\dagger$, $k k$, $\cdots$, must vanish.
This criterion imposes nontrivial constraint on the light-cone wave function $\varPhi_n(x)$:
\begin{align}
&\left(
\frac{m_{F}^{2}-2\lambda}{x}
+\frac{m_{B,r}^{2}-2\lambda}{1-x}
\right)\varPhi_{n}\left(x\right)-\lambda\pint_0^{1}\frac{dy}{\sqrt{(1-x)(1-y)}}\frac{2-x-y}{\left(x-y\right)^{2}}\varPhi_{n}\left(y\right)=M_{n}^{2}\varPhi_{n}\left(x\right).
\label{eq:hybrid:LC:12}
\end{align}
with the renormalized bosonic quark mass $m_{B,r}$ introduced in \eqref{mass:renormalization}.

Equation~\eqref{eq:hybrid:LC:12} is the desired BSE of ``baryon" in the IMF, which agrees
with what is originally derived by Aoki via diagrammatic approach~\cite{Aoki:1995dh}.

\subsection{BSEs of ``baryon" in FMF}

We proceed to derive the BSEs for the ``baryon" in hybrid $\QCDtwo$ in FMF,
based on the Hamiltonian approach in the context of the equal-time quantization.
To the best of our knowledge, this set of BSEs  has never been known before.

\subsubsection{Hamiltonian in the axial gauge}

Imposing the axial gauge $A^{a\,z}=0$ in \eqref{Lag:extended:t:Hooft:model}, the hybrid ${\rm QCD}_2$ lagrangian  reduces to
\begin{align}
  \mathcal{L}_{\text{hQCD}_2} &= \frac{1}{2}\left(\partial_{z}A_{0}^{a}\right)^{2}
    + (D_0\phi)^{\dagger}D_0\phi
    - (\partial_z\phi^\dagger)\partial_z\phi
    - m_B^{2}\phi^{\dagger}\phi
    + i\psi^{\dagger}\left(D_{0}+\gamma^{5}\partial_{z}\right)\psi-m_F\bar{\psi}\psi.
\end{align}
After Legendre transformation, we obtain
\begin{align}
  H &=\int dz\left[ \pi^{\dagger}\pi+|\partial_z\phi|^{2}+m_B^{2}|\phi|^{2}+\psi^{\dagger}(-i\gamma^{5}\partial_{z}+m_F\gamma^{0})\psi
    -\frac{\gs^2}{2} J^a\frac{1}{\partial_{z}^{2}}J^a \right],
\label{hQCD:ET:Hamiltonian}
\end{align}
with
\beq
  J^a = \psi^{\dagger}T^{a}\psi
    - i\left(\pi^{\dagger}T^{a}\phi-\phi^{\dagger}T^{a}\pi\right).
\eeq

\subsubsection{Dressed quark basis and Fermionization}

The bosonic quark field can be Fourier-expanded in terms of the annihilation and creation operators in accordance with \eqref{eq:scalar:ET:quant}.
The fermionic quark field is Fourier-expanded as~\cite{Jia:2018qee}
\begin{align}
\psi^{i}\left(z\right) & =\int\frac{dp}{2\pi}\frac{1}{\sqrt{2\widetilde{E}\left(p\right)}}
\left[b^{i}\left(p\right)u\left(p\right)+d^{i\dagger}\left(-p\right)v\left(-p\right)\right]e^{ipz},
\label{eq:eq1}
\end{align}
where the unregularized dispersion relation of the dressed fermionic quark ${\widetilde{E}(p)}$ is given by \cite{Jia:2018qee}
\beq
\widetilde{E}\left(p\right) = m_F\cos\theta\left(p\right)+p\sin\theta\left(p\right)+\frac{\lambda}{2}\int\frac{dk}
{\left(p-k\right)^{2}}\Theta\left(|k-p|-\rho\right)\cos[\theta\left(p\right)-\theta\left(k\right)],
\label{Bars:Green:dispersion:relation}
\\
\eeq
with $\theta(p)$ representing the Bogoliubov-chiral angle.
Note that the dispersion relation \label{Bars:Green:dispersion:relation} depends on an artificial IR regulator, which
can also be viewed as a gauge artifact,
since the energy of an colored object like QCD cannot be a physical quantity.

The Bogoliubov-chiral angle can be determined through the mass-gap equation in spinor $\QCDtwo$~\cite{Bars:1977ud}:
\begin{align}
p \cos\theta(p)-m_F \sin\theta(p) & =\frac{\lambda}{2} \pint_{-\infty}^{+\infty}\frac{dk}{\left(p-k\right)^{2}}
\sin\left[\theta\left(p\right)-\theta\left(k\right)\right].
\label{eq:spinor:mass_gap}
\end{align}
Like the mass gap equation in scalar $\QCDtwo$, \eqref{eq:scalar:mass_gap},
this mass gap equation can also be deduced via variational approach, {\it viz.},
by enforcing the vacuum energy to be minimized.

Analogous to the fermionization procedure employed in Sec.~\ref{Compound:Fermionization},
we introduce the following color-singlet compound operators:
\bseq
\begin{align}
B\left(p,q\right) & \equiv\sum_{i}b^{i\dagger}\left(p\right)b^{i}\left(q\right),
&
D\left(p,q\right) & \equiv\sum_{i}d^{i\dagger}\left(-p\right)d^{i}\left(-q\right),
\\
M\left(p,q\right) & \equiv\sum_{i}d_i (-p) b_i (q),
&
M^{\dagger}\left(p,q\right) & \equiv \sum_{i}b_i^{\dagger} (q) d_i^\dagger(-p),
\\
K\left(p,q\right) &  \equiv\frac{1}{\sqrt{\Nc}}\sum_{i}b^{i}\left(q\right)c^{i}\left(-p\right),
&
K^{\dagger}\left(p,q\right) & \equiv\frac{1}{\sqrt{\Nc}}\sum_{i}b^{i\dagger}\left(q\right)c^{i\dagger}\left(p\right),
\\
\overline{K}\left(p,q\right) & \equiv\frac{1}{\sqrt{\Nc}}\sum_{i}d^{i}\left(-p\right)a^{i}\left(q\right),
&
\overline{K}^{\dagger}\left(p,q\right) & \equiv\frac{1}{\sqrt{\Nc}}\sum_{i}d^{i\dagger}\left(-p\right)a^{i\dagger}\left(q\right).
\end{align}
\label{eq:lll}
\eseq

Note that the compound operator $K$ annihilates a fermionic quark and a bosonic antiquark,
and the compound operator $\overline{K}$ annihilates a fermionic antiquark and a bosonic quark.
The anticommutation relations among four fermionic compound color-singlet operators $K$,
$K^{\dagger}$, $\bar{K}$ and $\overline{K}^{\dagger}$ become
\bseq
\begin{align}
\left\{ K\left(k_{1},p_1\right),K^{\dagger}\left(k_{2},p_2\right)\right\}  & =\left(2\pi\right)^{2}\delta\left(k_{1}-k_{2}\right)\delta\left(p_1-p_2\right)+\mathcal{O}(1/N_c),
\\
\left\{ \overline{K}\left(k_{1}^{+},p_1^{+}\right),\overline{K}^{\dagger}\left(k_{2}^{+},p_2^{+}\right)\right\}  & =\left(2\pi\right)^{2}\delta\left(k_{1}-k_{2}\right)\delta\left(p_1-p_2\right)+\mathcal{O}(1/N_c),
\\
\left\{ K\left(k_{1},p_1\right),\overline{K}\left(k_{2},p_2\right)\right\}  & =\left\{ K^{\dagger}\left(k_{1},p_1\right),\overline{K}\left(k_{2},p_2\right)\right\} =0,
\\
\left\{ K\left(k_{1},p_1\right),\overline{K}^{\dagger}\left(k_{2},p_2\right)\right\}  & =\left\{ K^{\dagger}\left(k_{1},p_1\right),\overline{K}^{\dagger}\left(k_{2},p_2\right)\right\} =0.
\end{align}
\eseq

Since we are solely interested in the ``baryon" state,
the interaction term in \eqref{hQCD:ET:Hamiltonian} that couples
the bosonic quark sector with the fermionic sector is of pivotal importance.
It can be expressed in terms of the products of the
fermionic compound operators:
\begin{align}
& g_{s}\psi^{\dagger}T^{a}\psi\frac{1}{\partial_{z}^{2}}\left[ g_{s}i\left(\pi^{\dagger}T^{a}\phi-\phi^{\dagger}T^{a}\pi\right)\right]
\nn \\
&= -\pi\lambda\int\frac{dp_1}{2\pi}\int\frac{dp_2}{2\pi}\int\frac{dk_{1}}{2\pi}\int\frac{dk_{2}}{2\pi}
\frac{1}{\left(k_{2}-k_{1}\right)^{2}}
\Theta\left(|k_2-k_1|-\rho\right)2\pi\delta\left(p_2-p_1+k_2-k_1\right)\nn \\
&\qquad\times\Big\{f_{+}\left(k_{1},k_{2}\right)\cos\frac{\theta\left(p_1\right)-\theta\left(p_2\right)}{2}\left((K^{\dagger}\left(k_{2},p_1\right)K\left(k_{1},p_2\right)+\overline{K}^{\dagger}\left(p_2,k_{1}\right)\overline{K}\left(p_1,k_{2}\right)\right)
\nn \\
&\qquad +f_{-}\left(k_{1},k_{2}\right)\sin\frac{\theta\left(p_1\right)-\theta\left(p_2\right)}{2}\left(K^{\dagger}\left(k_{2},p_1\right)\overline{K}^{\dagger}\left(p_2,k_{1}\right)+\overline{K}\left(p_1,k_{2}\right)K\left(k_{1},p_2\right)\right)\Big\},
\end{align}
 where $f_\pm$ is given in \eqref{f:plus:minus}.

Expressing everything in terms of the color-singlet compound operators,
we then break the full Hamiltonian \eqref{hQCD:ET:Hamiltonian} into three pieces:
\beq
H =H_{0}+\colon H_{2}\colon+\colon H_{4}\colon+\mathcal{O}(1/\Nc)
\label{Breaking:Equal:Time:Hamiltonian:baryon}
\eeq
with
\bseq
\begin{align}
H_{0}= & N_c\int{dz}\bigg\{\int\frac{dk{(k^2+m_{1}^2)}}{2\pi\left(2E_k\right)}+\int\frac{dk{E_k}}{4\pi}   +\frac{\pi\lambda}{2}\int\frac{dk_1}{2\pi}\int\frac{dk_2}{2\pi}\frac{\left(E_{k_2}-E_{k_1}\right)^{2}}{\left(k_{1}-k_{2}\right)^{2}}\frac{1}{E_{k_1}E_{k_2}}
\\
& +\frac{dp_{1}}{2\pi}\mathrm{Tr}\left[\left(p_{1}\gamma^{5}+m_{2}\gamma^{0}\right)\varLambda_{-}\left(p_{1}\right)+\frac{\lambda}{2}\int\frac{dp_{2}}{\left(p_{2}-p_{1}\right)^2}
\Theta\left(|p_{2}-p_{1}|-\rho\right)\varLambda_{+}\left(p_{1}\right)\varLambda_{-}\left(p_{2}\right)\right]\bigg\},
\nn\\
:H_{2}: & = \int\frac{dk}{2\pi}\widetilde{\Pi}^{+}\left(k\right)\left(A\left(k\right)+C\left(k\right)\right)+\int \frac{dp}{2\pi}\widetilde{E}(p)\left(B\left(p,p\right)+D\left(p,p\right)\right),
\\
:H_{ 4}: & = -\pi\lambda\int\frac{dp_{1}}{2\pi}\int\frac{dp_2}{2\pi}\int\frac{dk_{1}}{2\pi}\int\frac{dk_{2}}{2\pi}\frac{1}{\left(k_{2}-k_{1}\right)^{2}} 
\Theta\left(|k_2-k_1|-\rho\right)2\pi\delta\left(p_2-p_1+k_2-k_1\right)
\nn\\
&\qquad\times\Big\{f_{+}\left(k_{1},k_{2}\right)\cos\frac{\theta\left(p_1\right)-\theta\left(p_2\right)}{2}\left((K^{\dagger}\left(k_{2},p_1\right)K\left(k_{1},p_2\right)+\overline{K}^{\dagger}\left(p_2,k_{1}\right)\overline{K}\left(p_1,k_{2}\right)\right)
\nn \\
&\qquad +f_{-}\left(k_{1},k_{2}\right)\sin\frac{\theta\left(p_1\right)-\theta\left(p_2\right)}{2}\left(K^{\dagger}\left(k_{2},p_1\right)\overline{K}^{\dagger}\left(p_2,k_{1}\right)+\overline{K}\left(p_1,k_{2}\right)K\left(k_{1},p_2\right)\right)\Big\},
\end{align}
\label{eq:h0}
\eseq
where
\beq
\varLambda_{\pm}\left(k\right)=T\left(k\right)\frac{1\pm\gamma^{0}}{2}T^{\dagger}\left(k\right),\qquad T\left(k\right)=\exp\left[-\frac{1}{2}\theta\left(k\right)\gamma^{z}\right].
\label{Lambda:and:T:def}
\eeq

Demanding that the $:H_{\text{2}}:$ piece to bear a diagonalized form separately for the dressed bosonic quarks and fermionic quarks,
we can obtain the mass-gap equations for both types of quarks,  \eqref{eq:scalar:mass_gap} and  \eqref{eq:spinor:mass_gap}.

\subsubsection{Bogoliubov transformation, diagonalization of Hamiltonian}

Like in the preceding subsection, the color confinement characteristic of QCD indicates that the
bosonic color-singlet compound operators $A$, $B$, $C$ and $D$ can not be independent,
yet at lowest order in $1/\Nc$ can be expressed in terms of the convolutions of the following fermionic color-singlet compound operators:
\bseq
\begin{align}
A\left(k_{1},k_{2}\right) & \rightarrow\intop\frac{dp}{2\pi}\overline{K}^{\dagger}\left(p,k_{1}\right)\overline{K}\left(p,k_{2}\right),\qquad
D\left(k_{1},k_{2}\right)\rightarrow\intop\frac{dp}{2\pi}\overline{K}^{\dagger}\left(k_{1},p\right)\overline{K}\left(k_{2},p\right),\\
C\left(k_{1},k_{2}\right) & \rightarrow\intop\frac{dp}{2\pi}K^{\dagger}\left(k_{1},p\right)K\left(k_{2},p\right),\qquad
B\left(k_{1},k_{2}\right)\rightarrow\intop\frac{dp}{2\pi}K^{\dagger}\left(p,k_{1}\right)K\left(p,k_{2}\right).
\end{align}
\eseq
Making these replacements in \eqref{Breaking:Equal:Time:Hamiltonian:baryon}, and only retaining the leading terms in
$1/\Nc$, the Hamiltonian now only depends on the fermionic compound operators $K$, $\overline{K}$ and their Hermitian conjugates:
\bseq
\begin{align}
:H_{2}: & =\iint\frac{dPdq}{\left(2\pi\right)^{2}}\left(\widetilde{\Pi}^{+}\left(P-q\right)+\widetilde{E}\left(q\right)\right)\overline{K}^{\dagger}\left(q,q-P\right)\overline{K}\left(q,q-P\right)
\nn \\
&+\left(\widetilde{\Pi}^{+}\left(P-q\right)+\widetilde{E}\left(q\right)\right)K^{\dagger}\left(q-P,q\right)K\left(q-P,q\right),\\
:H_{4}: & =-\frac{\lambda}{8\pi^{2}}\int dP\iint\frac{dqdk}{\left(k-q\right)^{2}}\Theta\left(|k-q|-\rho\right)
\big\{f_{+}\left(q-P,k-P\right)\cos\frac{\theta\left(k\right)-\theta\left(q\right)}{2} 
\nn\\
&\left(\overline{K}^{\dagger}\left(q,q-P\right)\overline{K}\left(k,k-P\right)
+K^{\dagger}\left(k-P,k\right)K\left(q-P,q\right)\right)
\nn\\
&+f_{-}\left(q-P,k-P\right)\sin\frac{\theta\left(k\right)-\theta\left(q\right)}{2}
\nn\\
&\left(\overline{K}\left(k,k-P\right)K\left(q-P,q\right)
+K^{\dagger}\left(k-P,k\right)\overline{K}^{\dagger}\left(q,q-P\right)\right)\big\}.
\end{align}
\eseq

To diagonalize the Hamiltonian, we invoke the Bogoliubov transformation \eqref{eq:1.11}, {\it viz.},
by expressing the fermionic color-singlet compound operators $K$, $\overline{K}$ in terms
of the annihilation/creation operators of the ``baryon" and ``anti-baryon":
\bseq
\begin{align}
K\left(q-P,q\right) & =\sqrt{\frac{2\pi}{|P|}}\sum_{n=0}^{\infty}\left[k_{n}\left(P\right)\varPhi_{n}^{+}\left(q,P\right)+\bar{k}_{n}^{\dagger}\left(-P\right)\overline{\varPhi}_{n}^{-}\left(q-P,-P\right)\right],\\
\overline{K}\left(q,q-P\right) & =\sqrt{\frac{2\pi}{|P|}}\sum_{n=0}^{\infty}\left[\bar{k}_{n}(-P)\overline{\varPhi}_{n}^{+}\left(q-P,-P\right)+k_{n}^{\dagger}\left(P\right)
\varPhi_{n}^{-}\left(q,P\right)\right].
\end{align}
\eseq
where $k_{n}$ annihilates the $n$-th excited ``baryon" state and $\bar{k}_n$ annihilates the $n$-th excited ``anti-baryon" state.
The Bogoliubov coefficient functions $\varPhi_{\pm}^{n}$ can be interpreted as the forward-moving/backward-moving wave functions of the
$n$-th excited ``baryon" state, whereas the Bogoliubov coefficient functions $\overline{\varPhi}_{\pm}^{n}$ can be interpreted as the
 forward-moving/backward-moving wave functions of the
$n$-th ``anti-baryon" state.

It is natural to anticipate that these ``baryon/anti-baryon" annihilation and creation operators
obey the standard anti-commutation relations:
\bseq
\begin{align}
\{ k_n(P),k_{m}^{\dagger}(P')\} & =2\pi\delta_{nm}\delta(P-P'), \qquad
\{\bar{k}_{n}(P),\bar{k}_{m}^{\dagger}(P')\} =2\pi\delta_{nm}\delta(P-P'),\\
\{k_{n}(P),\bar{k}_{m}(P')\} & =\{k_{n}^{\dagger}(P),\bar{k}_{m}^{\dagger}(P')\}=0.
\end{align}
\label{anticomu:relations:hybrid}
\eseq

The physical vacuum is defined by $k_{n}(P)|\Omega\rangle =\bar{k}_{n}(P)|\Omega\rangle  =0$ for any $P$ and $n$.
One then constructs a single ``baryon" and ``anti-baryon" states as
\beq
|P_n^{0},P\rangle  =\sqrt{2P_n^{0}}k_{n}^{\dagger}\left(P\right)|\Omega\rangle,
\qquad
|P_n^{0},P\rangle  =\sqrt{2P_n^{0}}\bar{k}_{n}^{\dagger}\left(P\right)|\Omega\rangle,
\label{Single:baryon:antibaryon:state}
\eeq
where $P_n^{0}=\sqrt{M_n^{2}+P^2}$ with $M_n$ denoting the mass of
the $n$-th ``baryon" state.

To be compatible with \eqref{anticomu:relations:hybrid}, the ``baryon" and ``anti-baryon" wave functions,
$\varPhi^{\pm}_n$ and $\overline{\varPhi}^{\pm}_{n}$,
must obey the following orthogonality and completeness conditions:
\bseq
\begin{align}
\int_{-\infty}^{+\infty}dp\left[\varPhi_{n}^{+}\left(p,P\right)\varPhi_{m}^{+}\left(p,P\right)+\varPhi_{n}^{-}\left(p,P\right)\varPhi_{m}^{-}\left(p,P\right)\right] & =|P|\delta^{nm},
\\
\int_{-\infty}^{+\infty}dp\left[\overline{\varPhi}_{n}^{-}\left(p-P,-P\right)\overline{\varPhi}_{m}^{-}\left(p-P,-P\right)+\overline{\varPhi}_{n}^{+}(p-P,-P)\overline{\varPhi}_{m}^{+}\left(p-P,-P\right)
\right] & =|P|\delta^{nm},
\\
\int_{-\infty}^{+\infty}dp\left[\varPhi_{n}^{+}\left(p,P\right)\overline{\varPhi}_{m}^{-}\left(p-P,-P\right)+
\varPhi_{n}^{-}\left(p,P\right)\overline{\varPhi}_{m}^{+}\left(p-P,-P\right)\right] & =0,
\\
\sum_{n=0}^{\infty}\left[\overline{\varPhi}_{n}^{+}\left(p-P,-P\right)\overline{\varPhi}_{n}^{+}\left(q-P,-P\right)+\varPhi_{n}^{-}(p,P)\varPhi_{n}^{-}\left(q,P\right)+
\right] & =|P|\delta\left(p-q\right),
\\
\sum_{n=0}^{\infty}\left[\varPhi_{n}^{+}\left(p,P\right)\varPhi_{n}^{+}\left(q,P\right)+\varPhi_{n}^{-}\left(p-P,-P\right)\varPhi_{n}^{-}\left(q-P,-P\right)\right] & =|P|\delta\left(p-q\right),
\\
\sum_{n=0}^{\infty}\left[\varPhi_{n}^{+}\left(p,P\right)\varPhi_{n}^{-}\left(q,P\right)+\overline{\varPhi}_{n}^{+}\left(q-P,-P\right)\overline{\varPhi}_{n}^{-}\left(p-P,-P\right)
\right] & =0.
\end{align}
\eseq
Switching to $k_{n}$ and $\bar{k}_{n}$ basis, we anticipate that all the non-diagonal terms in the
Hamiltonian in \eqref{Breaking:Equal:Time:Hamiltonian:baryon} vanish, and end up with the
desired diagonalized form:
\beq
H  = H_{0}^{'}+\int\frac{dP}{2\pi}\sum_{n} P_n^{0} \left[ k_{n}^{\dagger}(P) k_{n}(P)
+
\bar{k}_{n}^{\dagger}(P)\bar{k}_{n}(P)\right]+\mathcal{O} (1/\sqrt{N_c}).
\eeq

Demanding all the non-diagonalized terms to vanish, we find that the ``baryon" wave functions
$\varPhi_{\pm}^{n}$ must satisfy the following coupled integral equations:
\bseq
\begin{align}
  &\left(\Pi^{+}\left(P-p\right)+E\left(p\right)-P^{0}_n\right)\varPhi_n^{+}\left(p,P\right) \nn\\
  &=\frac{\lambda}{2}\pint\frac{dk}{\left(k-p\right)^{2}}
  \Big[f_{+}\left(k-P,p-P\right)\cos\frac{\theta\left(p\right)-\theta\left(k\right)}{2}\varPhi_n^{+}\left(k,P\right)\nn \\
  &\qquad+f_{-}\left(k-P,p-P\right)\sin\frac{\theta\left(p\right)-\theta\left(k\right)}{2}\varPhi_n^{-}\left(k,P\right)\Big],
    \label{eq:ds3}
  \\
&\left(\Pi^{+}\left(P-p\right)+E\left(p\right)+P^{0}_n\right)\varPhi_n^{-}\left(p,P\right) \nn\\
  &=\frac{\lambda}{2}\pint\frac{dk}{\left(k-p\right)^{2}}
    \Big[f_{+}\left(k-P,p-P\right)\cos\frac{\theta\left(p\right)-\theta\left(k\right)}{2}\varPhi_n^{-}\left(k,P\right) \nn \\
  &\qquad-f_{-}\left(k-P,p-P\right)\sin\frac{\theta\left(p\right)-\theta\left(k\right)}{2}\varPhi_n^{+}\left(k,P\right)\Big],
    \label{eq:ds4}
\end{align}
\label{eq:baryon:BSE:FMF}
\eseq
where the regularized dressed quark energy $E(p)$ is defined by~\cite{Bars:1977ud}:
\begin{align}
E\left(p\right) &= \widetilde{E}\left(p\right)-\frac{\lambda}{\rho}
= m_F\cos\theta\left(p\right)+p\sin\theta\left(p\right)+\frac{\lambda}{2}\pint\frac{dk}{\left(p-k\right)^{2}} \cos[\theta\left(p\right)-\theta\left(k\right)].
\end{align}

Equations~\eqref{eq:baryon:BSE:FMF} 
represent the BSEs for a ``baryon" 
in hybrid $\QCDtwo$ in FMF,
which constitute the main new results of this paper.

When boosted to the IMF, one can show that the BSEs \eqref{eq:baryon:BSE:FMF} reduce to its LF counterpart, \eqref{eq:hybrid:LC:12}.
In the other words, as the ``baryon" momentum is increasing, the backward-moving wave function
$\varPhi_{-}^{n}$ quickly fades away, while the forward-moving wave function $\varPhi_{+}^{n}$ approaches the light-front
wave function $\varPhi^{n}(x)$.

\section{Numerical results}
\label{sec:numerical}

\begin{figure}[!htbp]
  \includegraphics[scale=0.5]{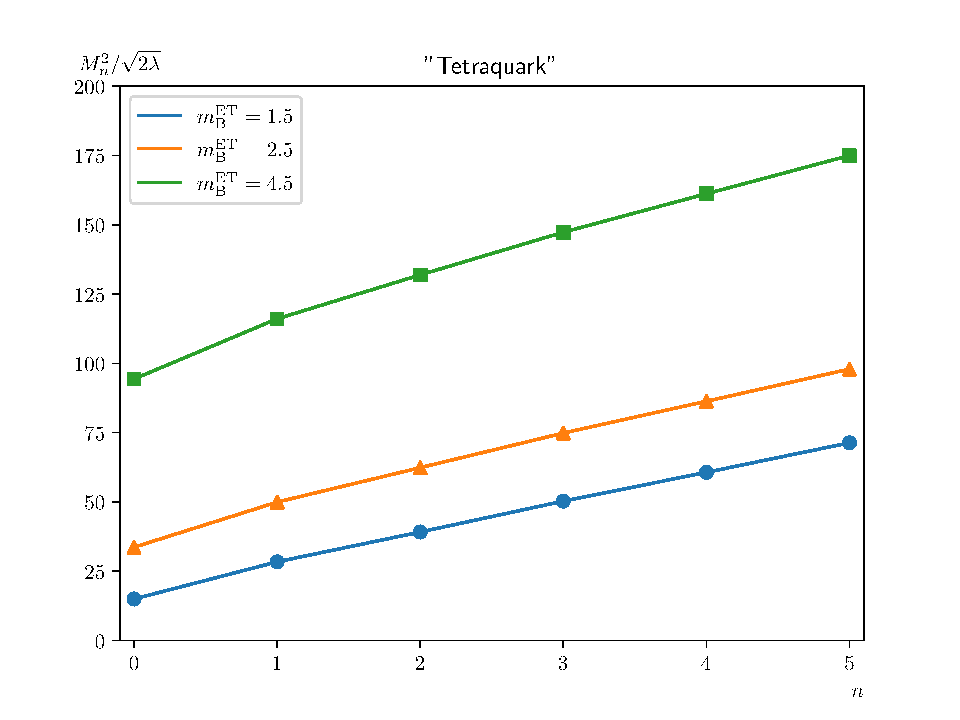} \\
  \includegraphics[scale=0.5]{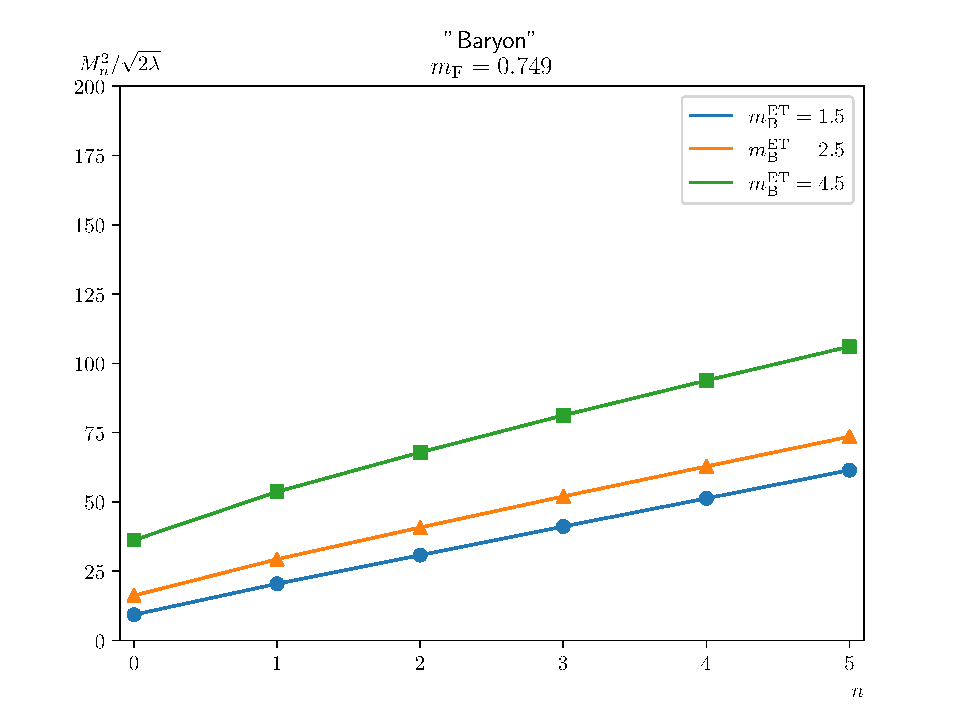}
  \includegraphics[scale=0.5]{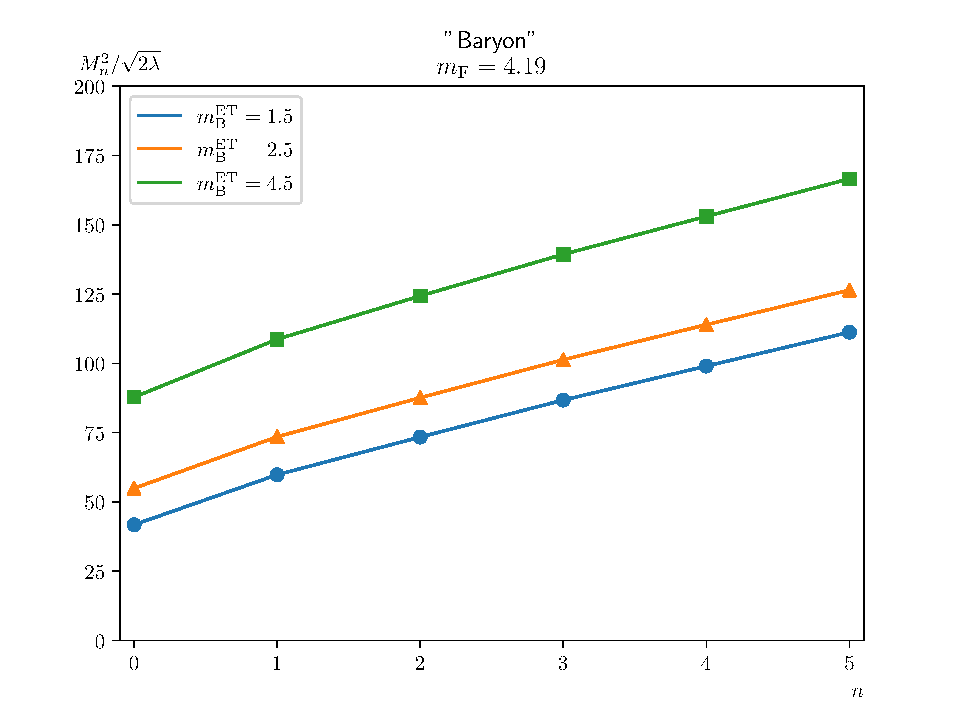}
  \caption{Mass spectra of the ``tetraquark" and ``baryon" with some different set of quark masses.}
  \label{eq:fig1}
\end{figure}

\begin{figure}
  \includegraphics[scale=0.5]{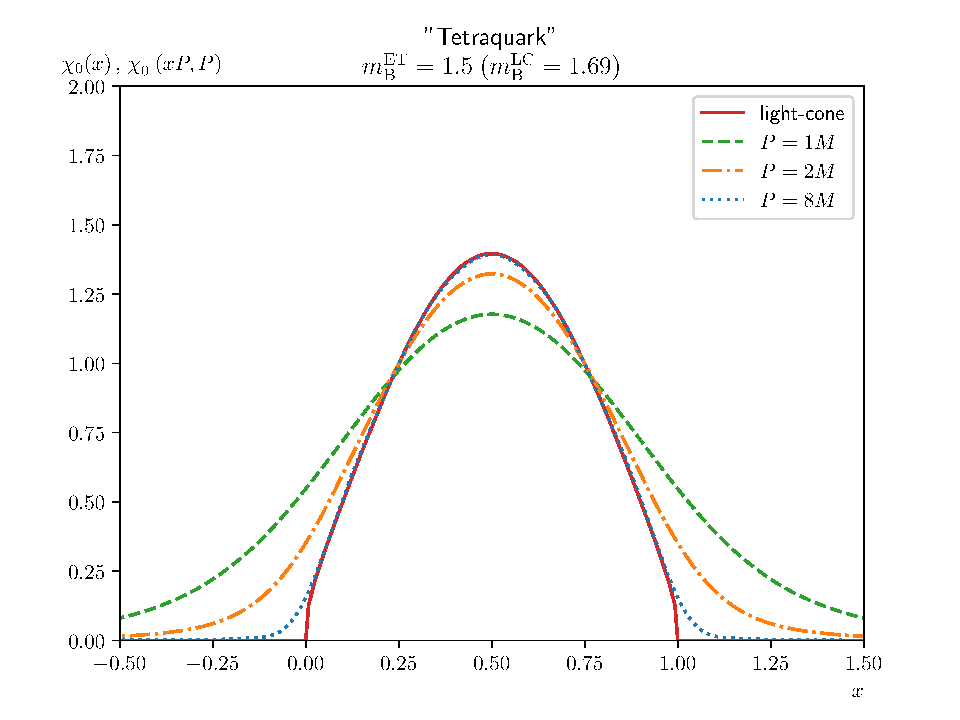}
  \includegraphics[scale=0.5]{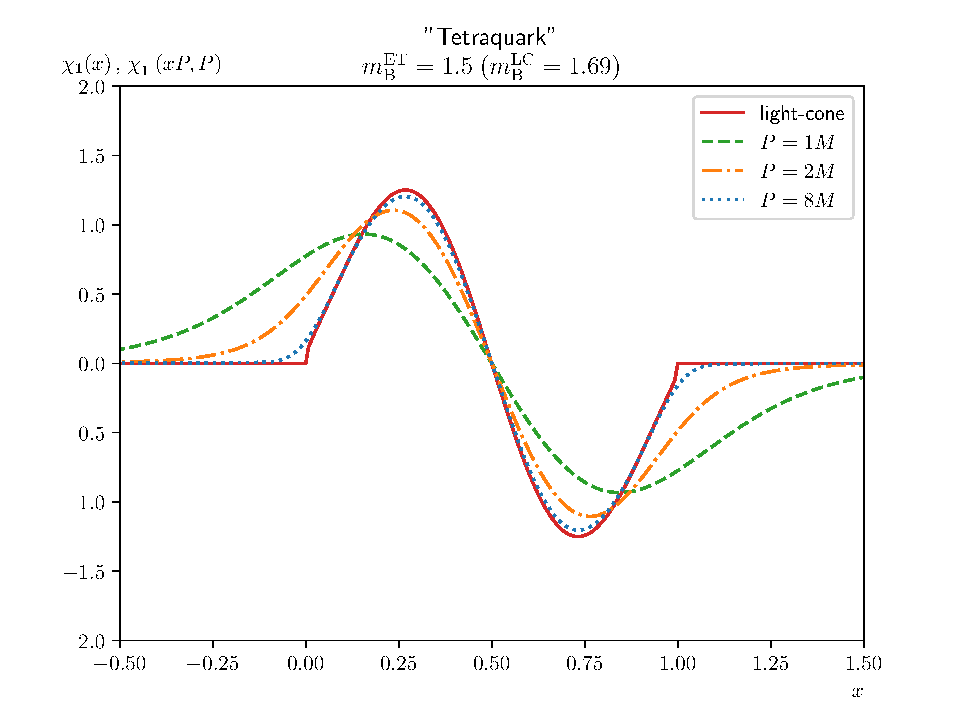}\\
  \includegraphics[scale=0.5]{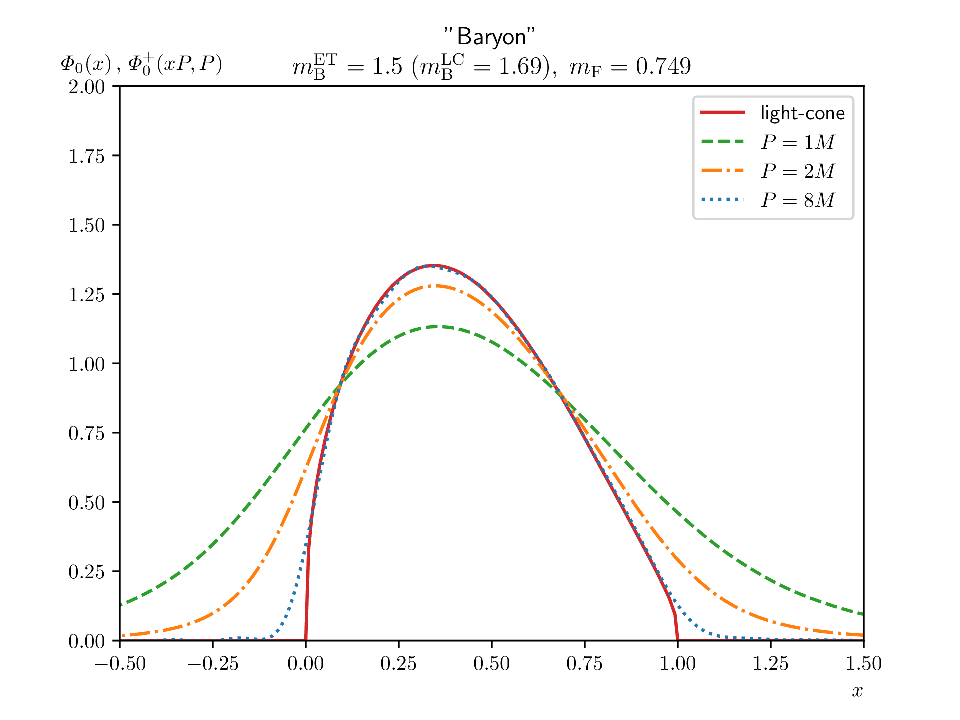}
  \includegraphics[scale=0.5]{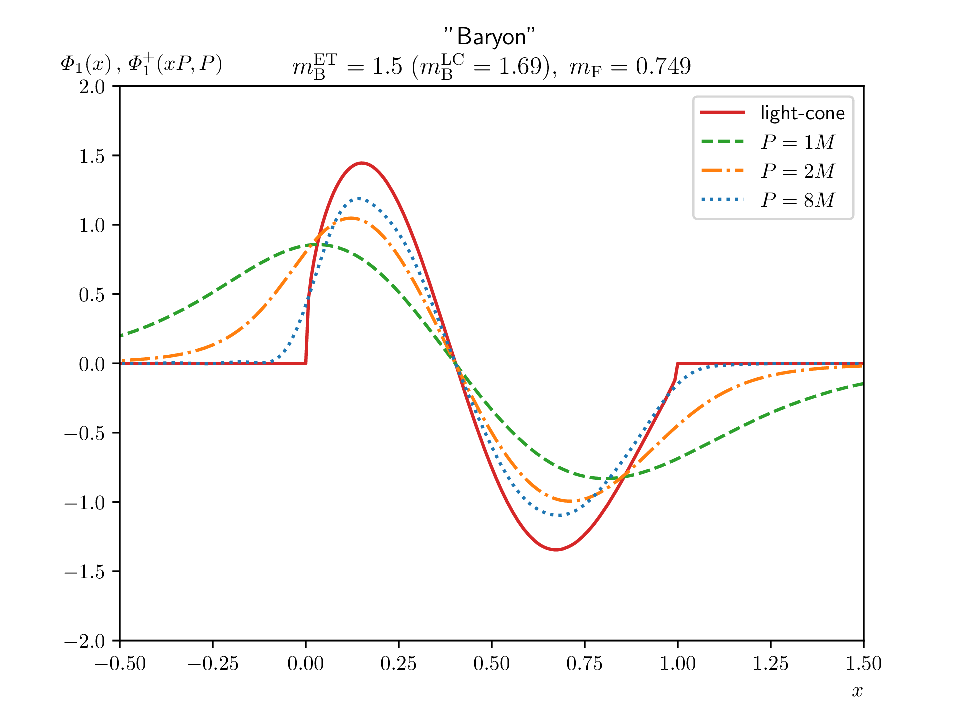}\\
  \includegraphics[scale=0.5]{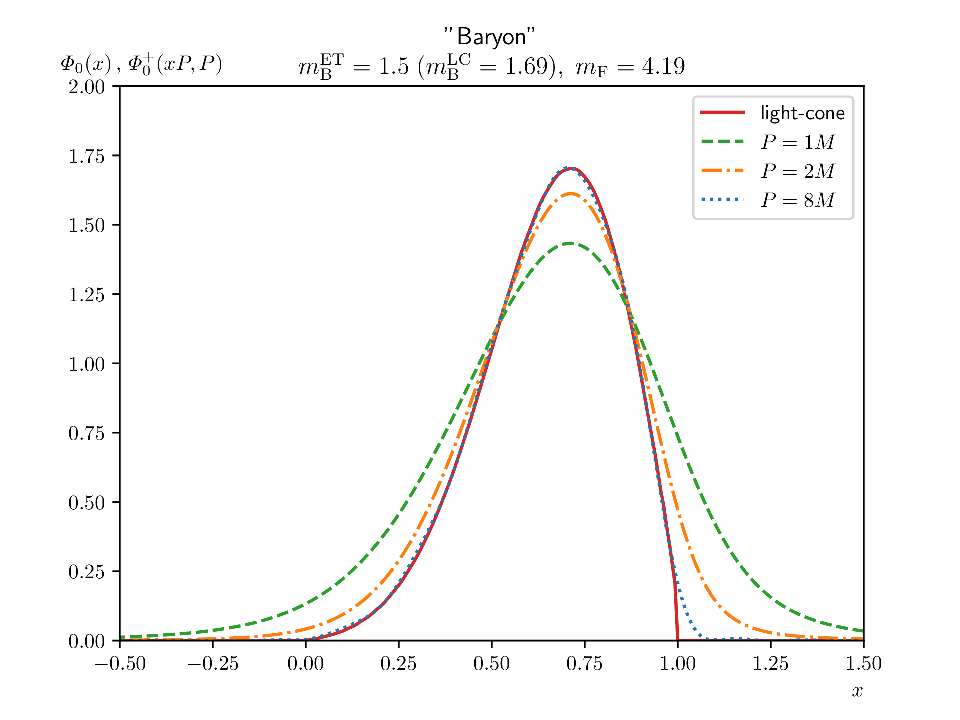}
  \includegraphics[scale=0.5]{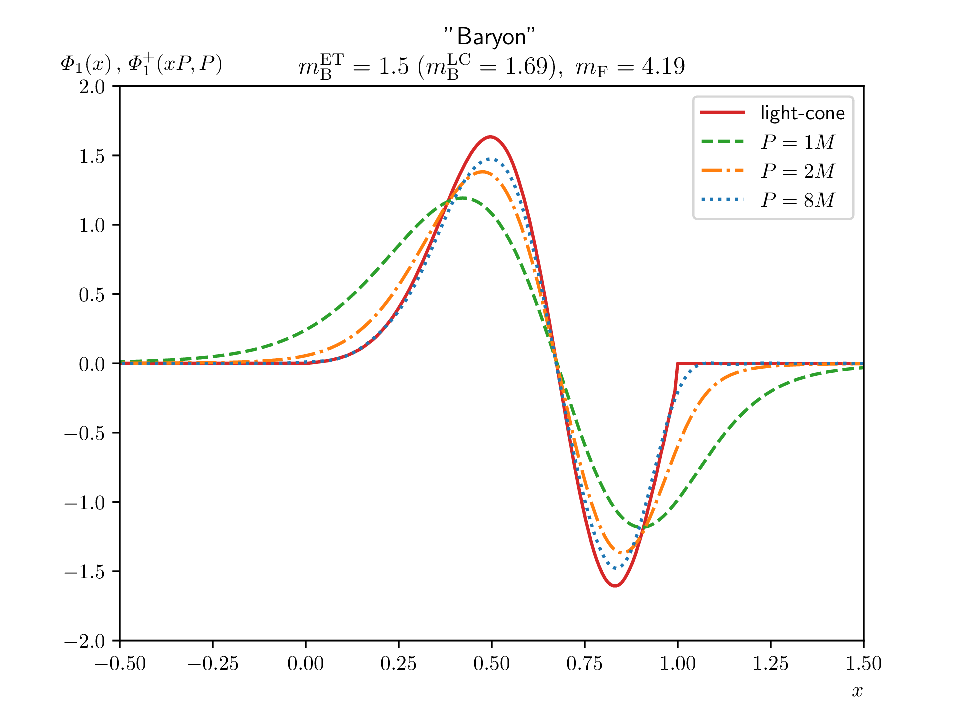}
  \caption{Profiles of the forward-moving components of the wave functions of ``tetraquark" and ``baryons", viewed from different finite momentum frames.
     The wave functions of the ground state are shown in the left column, while those of the first excited state are shown in the right column.
The solid curves represent the corresponding light-cone wave functions. }
  \label{eq:fig2}
\end{figure}

The numerical recipe of solving the BSEs in 't Hooft model has been adequately discussed in literature.
We follow the approach outlined in \cite{Brower:1978wm} to solve the BSEs of the ``tetraquark" and ``baryon" in IMF, and follow the approach based on
Hermite function expansion~\cite{Li:1987hx,Jia:2017uul} to solve the BSEs of the ``extotic" hadrons in FMF.

The dimensionful 't Hooft coupling is usually taken to be $\sqrt{2\lambda}=340\text{ MeV}$, which is close to the value of string tension in $\text{QCD}_4$.
For simplicity, we use $\sqrt{2\lambda}$ as the unit of the mass in the rest of this section.
The renormalized mass of bosonic quark in equal-time quantization is taken to be $m_B^\text{ET}\equiv m_{B,\tilde{r}}=1.5$, which is equivalent to the one
defined in LF quantization, $m_\text{B}^\text{LC}\equiv m_{B,r}=1.69$~\footnote{We mention that the connection of two different renormalized bosonic quark masses
is given in (B7), since we follow \cite{Ji:2018waw} to impose the bosonic quark renormalization condition in accordance with \eqref{quark:mass:counterterm:sQCD:ETQ}.
Of course, we could have chosen a different renormalization scheme as specified in (B8), such that $m_{B,\tilde{r}}=m_{B,r}$.}.
As for fermionic quarks, we choose the strange quark mass as $m_F=0.749$ and
charm quark mass as $m_F=4.19$~\cite{Jia:2017uul}.

In Fig.~\ref{eq:fig1}, the mass spectra of ``tetraquark" and ``baryon"  with different
quark species are plotted against the principal quantum number $n$. One can observe the tendency that the squared hadron mass linearly grows with $n$ when
the principle quantum number gets large. This pattern is identical with the Regge trajectory observed in the original 't Hooft model,
which is also somewhat analogous to the Regge trajectories observed in the real world, where the squared mass of the excited hadronic state
linearly grows with the spin.

In passing we mention a technical nuisance, {\it i.e.}, it is found that, when numerically solving the Shei-Tsao equation,
the renormalized bosonic quark mass $m_\text{B}^\text{LC}$ can not be chosen less than unity.
As pointed out in \cite{Visnjic:1995us}, such case may corresponds to a strongly coupled regime,
where the the LF wave function near the endpoints is no longer real-valued.

In Fig~\ref{eq:fig2}, we plot the equal-time and light-front (LF) meson wave functions pertaining to ``tetraquark" and ``baryon", including
both the ground states and the first excited states. For a ``tetraquark" consisting of a single flavor of bosonic quark,
the LF wave functions with even/odd $n$ are symmetric/antisymmetric under the exchange $x \leftrightarrow (1-x)$ due to the charge conjugation symmetry.
The LF wave functions always vanish in the endpoints $x = 0,1$.

It is interesting to see how the bound-state wave functions obtained in equal-time quantization evolve with the hadron's momentum.
As is evident in Fig~\ref{eq:fig2}, as the hadron momentum increases, the forward-moving wave functions of ``tetraquark", $\chi_{+}$, and of ``baryon",
$\varPhi_{+}$, rapidly converge to the corresponding light-cone wave functions. We have also numerically verified that, the backward-moving wave functions,
$\chi_{-}$ and $\varPhi_{-}$, do quickly fade away with increasing hadron momentum. This finding is identical with what is observed in the numerical study of
the original 't Hooft model~\cite{Jia:2017uul}, which is compatible with the tenet of LaMET.

\section{Summary
\label{sec:summary}}

In this work, we have made a comprehensive study of the extended 't Hooft model, {\it viz.},
two-dimensional QCD including both fermionic and bosonic quarks in $N_c\to \infty$ limit.
We focus on derivations of the bound-state equations pertaining to two types of hadrons,
a ``tetraquark" composed of a bosonic quark and a bosonic antiquark,
and a ``baryon" composed of a fermionic quark and bosonic antiquark.
Using the Hamiltonian approach, we derive these BSEs from the perspectives of light-front and
equal-time quantization, which are associated with the IMF and FMF, respectively.
We confirm the known results, such as those BSEs of ``tetraquark" in IMF~\cite{Shei:1977ci} and in FMF~\cite{Ji:2018waw}.
We have paid special attention to the issue concerning quark mass renormalization.  It is found that,
the renormalized bosonic quark mass  in scalar $\QCDtwo$ in light-front quantization may not coincide with that in equal-time quantization,
and the relationship between these two types of renormalized quark mass is established. Moreover, for the first time we also present a diagrammatic derivation of the ``tetraquark" BSEs
in FMF.

We have also confirmed the BSE of ``baryon" in the extended 't Hooft model in IMF~\cite{Aoki:1993ma}.
The main new result of this work is to derive, for the first time, the BSEs of ``baryon" in FMF in the context of equal-time quantization.

We have also conducted a comprehensive numerical study of mass spectra of ``tetraquark" and ``baryon".
The Regge trajectories are explicitly demonstrated. We have also obtained the profiles of the wave functions of the ground and first excited states of ``tetraquark" and ``baryon",
viewed from different FMFs. We have numerically verified that,  when the ``tetraquark" and ``baryon" are boosted to IMF,
the forward-moving components of the bound-state wave functions approach the corresponding light-cone wave functions,
and the backward-moving components fade away.

\begin{acknowledgments}
This work was supported in part by the National Science
Foundation of China under the Grants No.~12475090,11925506, 12435004, and the
Natural Science Foundation of Shandong province under the
Grant No. ZR2022ZD26. The work of Z. M. is also supported in part by the National Natural Science Foundation of China No. 12347145, No. 12347105.
\end{acknowledgments}

\appendix

\section{{Diagrammatic derivation of the bound-state equation for scalar $\text{QCD}_2$ in equal-time quantization}
\label{Diagram:derivation:sQCD2}}

In this Appendix, we employ the diagrammatic technique to rederive the bound-state equations of ``tetraquark"
in scalar ${\rm QCD}_2$ in the context of equal-time quantization. 

\subsection{Feynman Rules}

To quantize scalar ${\rm QCD}_2$ in equal-time, we adopt the axial gauge $A^z = 0$ and proceed by
expanding the Lagrangian as follows:
\begin{align}
  \mathcal{L}_{\text{sQCD}_2} &=
    \tr\left(\partial_z A^0\right)^2
    + \left(\partial^\mu\phi^\dagger\right)\partial_\mu\phi
    - m^2\phi^{\dagger}\phi \nn\\
  &\qquad -i\gs\left(\partial_0\phi^\dagger\right) A^0\phi
    + i\gs\phi^\dagger A^0\partial_0\phi
    + \gs^2\phi^\dagger A^0A^0\phi.
\label{eq:Lag:sQCD2:ex}
\end{align}
The Feynman rules can be directly derived from \eqref{eq:Lag:sQCD2:ex} and are summarized in
Tab. \ref{tab:sQCD:Feyn:rules}. A key challenge in equal-time quantization of scalar ${\rm QCD}_2$,
compared to its spinor counterpart, lies in the presence of seagull vertices. These vertices
disrupt the rainbow-ladder topology even at leading order in the $1/\Nc$ expansion.
\begin{table}[!hbtp]
  \centering
  \caption{Original Feynman rules for sQCD$_{2}$ in the axial gauge}
  \label{tab:sQCD:Feyn:rules}
  \begin{tabular}{ccc}
  \hline
     Building blocks & Double lines & Feynman rules \\
  \hline
     \\

     \begin{tikzpicture}[baseline]
      \coordinate (I) at (-1, 0);
      \coordinate (O) at (1, 0);
      \draw[gluon=3.9] (I)
         -- node[pos=0.5, above]{$k$}
         (O);
     \end{tikzpicture} &
     \hspace{-0.9em}\loosebox{\begin{tikzpicture}[baseline]
      \coordinate (I1) at (-1, 0.5);
      \coordinate (I2) at (-1, 0.7);
      \coordinate (O1) at (1, 0.5);
      \coordinate (O2) at (1, 0.7);
      \draw[dbline] (I1) -- (O1);
      \draw[dbline] (O2) -- (I2);

      \coordinate (I3) at (-1, -0.5);
      \coordinate (I4) at (-1, -0.3);
      \coordinate (O3) at (1, -0.5);
      \coordinate (O4) at (1, -0.3);
      \coordinate (V1) at (-0.1, -0.5);
      \coordinate (V2) at (-0.1, -0.3);
      \coordinate (V3) at (0.1, -0.5);
      \coordinate (V4) at (0.1, -0.3);
      \draw (-1, -0.4) node[left] {$-\frac{1}{\Nc}$};
      \draw[dbline] (I3) -- (V1); \draw (V1) -- (V2); \draw[dbline] (V2) -- (I4);
      \draw[dbline] (O4) -- (V4); \draw (V4) -- (V3); \draw[dbline] (V3) -- (O3);
     \end{tikzpicture}}\hspace{1.6em} &
     $\displaystyle{\frac{i}{\left(k^z\right)^2}}$ \\
     \\

     \adjustbox{padding = 0 2ex 0 0}{\begin{tikzpicture}[baseline]
      \coordinate (I) at (-1, 0);
      \coordinate (O) at (1, 0);
      \draw[dashed, fermion] (I) -- node[pos=0.5, above]{$p$} (O);
     \end{tikzpicture}} &
     \begin{tikzpicture}[baseline]
      \coordinate (I1) at (-1, 0);
      \coordinate (O1) at (1, 0);
      \draw[dbline] (I1) -- (O1);
     \end{tikzpicture} &
     \loosebox{$\displaystyle{\frac{i}{p^2 - m^2 + i\epsilon}}$} \\
     \\

    \loosebox{\begin{tikzpicture}[baseline=(current bounding box.center)]
      \coordinate (I) at (-1, -0.2);
      \coordinate (O) at (1, -0.2);
      \coordinate (V) at (0, 0);
      \coordinate (O1) at (0, 1.06);

      \draw[dashed, fermion] (I) node[below]{$p$} -- (V);
      \draw[dashed, fermion] (V) -- (O) node[below]{$p'$};
      \draw[gluon] (V)  -- (O1);
     \end{tikzpicture}} &
     \begin{tikzpicture}[baseline=(current bounding box.center)]
      \draw[dbline] (-1.1, -0.2) -- (-0.1, 0);
      \draw[dbline] (-0.1, 0) -- (-0.1, 1.06);
      \draw[dbline] (0.1, 1.06) -- (0.1, 0);
      \draw[dbline] (0.1, 0) -- (1.1, -0.2);
     \end{tikzpicture} &
     $\displaystyle{\frac{i\gs}{\sqrt{2}}\left(p^0 + p^{\prime0}\right)}$ \\

     \loosebox{\begin{tikzpicture}[baseline=(current bounding box.center)]
      \coordinate (I) at (-1, -0.2);
      \coordinate (O) at (1, -0.2);
      \coordinate (V) at (0, 0);
      \coordinate (O1) at (-1, 0.95);
      \coordinate (O2) at (1, 0.95);

      \draw[dashed, fermion] (I) -- (V);
      \draw[dashed, fermion] (V) -- (O);
      \draw[gluon=4.1] (V)  -- (O1);
      \draw[gluon=4.1] (V)  -- (O2);
     \end{tikzpicture}} &
     \loosebox{\specialcell{
      \begin{tikzpicture}
         \draw[dbline] (-1.14, -0.2) -- (-0.14, 0);
         \draw[dbline] (-0.14, 0) -- (-1.14, 1) node[above left]{$1$};
         \draw[dbline] (-1, 1.14) -- (0, 0.14);
         \draw[dbline] (0, 0.14) -- (1, 1.14);
         \draw[dbline] (1.14, 1) node[above right]{$2$} -- (0.14, 0);
         \draw[dbline] (0.14, 0) -- (1.14, -0.2);
      \end{tikzpicture} \\
      $+ \left(1\leftrightarrow 2\right)$
      }} &
      $\displaystyle{\frac{i\gs^2}{2}}$ \\
     \hline
  \end{tabular}
\end{table}

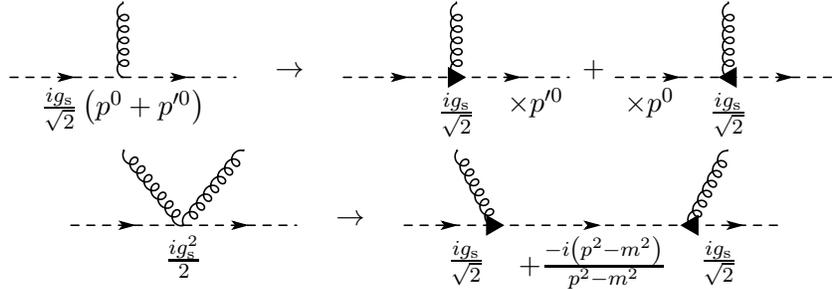
\begin{figure}[!hbtp]
  \centering
  \begin{tikzpicture}[baseline]
     \coordinate (I) at (-1.5, 0);
     \coordinate (V) at (0, 0);
     \coordinate (O) at (1.5, 0);
     \coordinate (G) at (0, 1);

     \draw[dashed, fermion] (I) -- (V); \draw[dashed, fermion] (V) -- (O);
     \draw[gluon=4.3] (V) -- (G);
     \node[below] at (V) {$\frac{i\gs}{\sqrt{2}}\left(p^0+p^{\prime0}\right)$};
  \end{tikzpicture} $\quad\to\quad$
  \begin{tikzpicture}[baseline]
     \coordinate (I) at (-1.5, 0);
     \coordinate (V) at (0, 0);
     \coordinate (O) at (1.5, 0);
     \coordinate (G) at (0, 1);

     \draw[dashed, fermion] (I) -- (V);
     \draw[dashed, fermion] (V) -- (O) node[pos=0.7, below] {$\frac{i p^{\prime0}}{p^2-m^2}$};
     \draw[gluon=4.3] (V) -- (G);
     \node[trinode] at (V) {}; \node[below, yshift=-2] at (V) {$\frac{i\gs}{\sqrt{2}}$};
  \end{tikzpicture} $+$
  \begin{tikzpicture}[baseline]
     \coordinate (I) at (-1.5, 0);
     \coordinate (V) at (0, 0);
     \coordinate (O) at (1.5, 0);
    \coordinate (G) at (0, 1);

     \draw[dashed, fermion] (I) -- (V) node[pos=0.3, below] {$\frac{i p^{0}}{p^2-m^2}$};
     \draw[dashed, fermion] (V) -- (O);
     \draw[gluon=4.3] (V) -- (G);
     \node[rtrinode] at (V) {}; \node[below, yshift=-2] at (V) {$\frac{i\gs}{\sqrt{2}}$};
  \end{tikzpicture}\\

  \begin{tikzpicture}[baseline]
     \coordinate (I) at (-1.5, 0);
    \coordinate (V) at (0, 0);
     \coordinate (O) at (1.5, 0);
     \coordinate (G1) at (-0.8, 1);
     \coordinate (G2) at (0.8, 1);

     \draw[dashed, fermion] (I) -- (V); \draw[dashed, fermion] (V) -- (O);
     \draw[gluon=4.25] (V) -- (G1); \draw[gluon=4.25] (V) -- (G2);
     \node[below] at (V) {$\frac{i\gs^2}{2}$};
  \end{tikzpicture} $\quad\to\quad$
  \begin{tikzpicture}[baseline]
     \coordinate (I) at (-2.5, 0);
     \coordinate (V1) at (-1.3, 0);
     \coordinate (V2) at (1.3, 0);
     \coordinate (O) at (2.5, 0);
     \coordinate (G1) at (-1.8, 1);
     \coordinate (G2) at (1.8, 1);
     \draw[dashed, fermion] (I) -- (V1);
     \draw[dashed, fermion] (V1) -- (V2)
      node[pos=0.5, below] {$+\frac{-i\left(p^2-m^2\right)}{p^2-m^2}$};
     \draw[dashed, fermion] (V2) -- (O);
     \draw[gluon=4.2] (V1) -- (G1); \draw[gluon=4.2] (V2) -- (G2);
     \node[trinode] at (V1) {}; \node[below left, yshift=-2] at (V1) {$\frac{i\gs}{\sqrt{2}}$};
     \node[rtrinode] at (V2) {}; \node[below right, yshift=-2] at (V2) {$\frac{i\gs}{\sqrt{2}}$};
  \end{tikzpicture}
  \caption{Modification of the Feynman rules to remove the seagull vertex. }
  \label{fig:sQCD:seagull:mod}
\end{figure}

The solution proceeds in a straightforward manner: we decompose each seagull vertex into two
$q\bar{q}g$ vertices, and reorganize its contribution into a new term of the quark propagator between
these two vertices. However, this cannot be done naively, since the Feynman rule for the $q\bar{q}g$
vertex depends on the momenta of the two adjacent quark lines, while a term in the quark propagator
only accounts for the momentum flowing through it. To address this, we must also split the $q\bar{q}g$
vertex into different types and distribute the quark momenta into the two adjacent quark propagators
separately. This process is illustrated in Fig. \ref{fig:sQCD:seagull:mod}.
To fully absorb the seagull vertex, it becomes necessary to introduce four distinct types of quark
propagators, which are shown in Fig. \ref{fig:sQCD:prop:mod}.
In retrospect, these propagators emerge quite naturally, and can, to some extent, be associated with
the propagators from $\phi^\dagger\left(y\right), -i\pi^\dagger\left(y\right)$ to
$\phi\left(x\right), i\pi\left(x\right)$ within the Hamiltonian formalism.
The situation here is somewhat analogous though not identical to the relationship between
canonical quantization and path integral quantization for the massive vector field.
The contact term arising from the seagull vertex cancels out the $\left(p^0\right)^2$ numerator
in the quark propagator, yielding the correct residue for $p^0$. For convenience, we define
$\omega_p^2 \equiv \left(p^z\right)^2 + m^2 -i \epsilon$.

\begin{figure}[!hbtp]
  \centering
  \begin{tikzpicture}[baseline]
     \coordinate (I) at (-1, 0);
     \coordinate (O) at (1, 0);

     \draw[dashed, fermion] (I) -- (O);
     \node[rtrinode] at (I) {}; \node[trinode] at (O) {};
     \node[below, yshift=-4] at (V)
      {$\scriptstyle{\bra{0}\T\phi\left(x\right)\phi^\dagger\left(y\right)\ket{0}}$};
  \end{tikzpicture} $\quad\displaystyle{=\frac{i}{p^2-m^2+i\epsilon}}$ \\

  \begin{tikzpicture}[baseline]
     \coordinate (I) at (-1, 0);
     \coordinate (O) at (1, 0);

     \draw[dashed, fermion] (I) -- (O);
     \node[rtrinode] at (I) {}; \node[rtrinode] at (O) {};
     \node[below, yshift=-4] at (V)
      {$\scriptstyle{\bra{0}\T i\pi\left(x\right)\phi^\dagger\left(y\right)\ket{0}}$};
  \end{tikzpicture} $\quad=\quad$
  \begin{tikzpicture}[baseline]
     \coordinate (I) at (-1, 0);
     \coordinate (O) at (1, 0);

     \draw[dashed, fermion] (I) -- (O);
     \node[trinode] at (I) {}; \node[trinode] at (O) {};
     \node[below, yshift=-4] at (V)
      {$\scriptstyle{\bra{0}\T\phi\left(x\right)\left(-i\pi^\dagger\left(y\right)\right)\ket{0}}$};
  \end{tikzpicture} $\quad\displaystyle{=\frac{ip^0}{p^2-m^2+i\epsilon}}$ \\

  \begin{tikzpicture}[baseline]
     \coordinate (I) at (-1, 0);
     \coordinate (O) at (1, 0);

     \draw[dashed, fermion] (I) -- (O);
     \node[trinode] at (I) {}; \node[rtrinode] at (O) {};
     \node[below, yshift=-4] at (V)
      {$\scriptstyle{\bra{0}\T\pi\left(x\right)\pi^\dagger\left(y\right)\ket{0}}$};
  \end{tikzpicture} $\quad\displaystyle{
     = \frac{i\left(p^0\right)^2 - i\left(p^2-m^2\right)}{p^2-m^2+i\epsilon}
     = \frac{i\left[\left(p^z\right)^2+m^2\right]}{p^2-m^2+i\epsilon}
  }$
  \caption{Modification of the quark propagators. }
  \label{fig:sQCD:prop:mod}
\end{figure}

To streamline the calculation, we express the four quark propagators (Fig. \ref{fig:sQCD:prop:mod})
in the form of a $2\times2$ matrix:
\begin{equation}
  D^{(0)}\left(p\right) = \frac{i\left(p^{0}\sigma_1 + \mathcal{P}_{+} +
  \omega_p^2 \mathcal{P}_{-}\right)}{p^2-m^2+i\epsilon},
\label{eq:D:free}
\end{equation}
where the projectors $\mathcal{P}_\pm \equiv \left(1\pm\sigma_3\right)/2$ satisfy the orthogonality
and completeness relations:
\begin{equation}
  \mathcal{P}_\pm^2 = \mathcal{P}_\pm,\quad \mathcal{P}_\pm\mathcal{P}\mp = 0,\quad \mathcal{P}_{+} + \mathcal{P}_{-} = 1,
\end{equation}
and the algebra:
\begin{equation}
  \sigma_1\mathcal{P}_\pm = \mathcal{P}_\mp \sigma_1,\quad
  \left(X\sigma_1 + Y\mathcal{P}_{+} + Z \mathcal{P}_{-}\right)\left(X\sigma_1 - Z\mathcal{P}_{+} - Y\mathcal{P}_{-}\right)
    = X^2-YZ,
\end{equation}
for scalar quantities $X,Y,Z$. It is important to emphasize that the introduction of the
Pauli $\sigma$-matrices here is unrelated to spinors. In fact, the mass dimensions of the matrix
elements in \eqref{eq:D:free} are not homogeneous. However, this particular choice of matrix
multiplication ensures that only terms of uniform dimension will be added within the matrix elements.
The modified Feynman rules are summarized in Tab. \ref{tab:sQCD:Feyn:rules:mod}.

\begin{table}[!hbtp]
  \centering
  \caption{Modified Feynman rules for sQCD$_2$ in the axial gauge. }
  \label{tab:sQCD:Feyn:rules:mod}
  \begin{tabular}{cc}
  \hline
     Building blocks & Feynman rules \\
  \hline
  \\

     \begin{tikzpicture}[baseline]
      \coordinate (I) at (-1, 0);
      \coordinate (O) at (1, 0);
      \draw[gluon=3.9] (I)
         -- node[pos=0.5, above]{$k$}
         (O);
     \end{tikzpicture} &
     $\displaystyle{\frac{i}{\left(k^z\right)^2}}$ \\
     \\

     \adjustbox{padding = 0 2ex 0 0}{\begin{tikzpicture}[baseline]
      \coordinate (I) at (-1, 0);
      \coordinate (O) at (1, 0);
      \draw[dashed, fermion] (I) -- node[pos=0.5, above]{$p$} (O);
     \end{tikzpicture}} &
     \loosebox{$\displaystyle{\frac{i}{p^0\sigma_1 - \omega_p^2\mathcal{P}_{+} -\mathcal{P}_{-}}}$} \\
     \\

     \loosebox{\begin{tikzpicture}[baseline=(current bounding box.center)]
      \coordinate (I) at (-1, -0.2);
      \coordinate (O) at (1, -0.2);
      \coordinate (V) at (0, 0);
      \coordinate (O1) at (0, 1.06);

      \draw[dashed, fermion] (I) node[below]{$p$} -- (V);
      \draw[dashed, fermion] (V) -- (O) node[below]{$p'$};
      \draw[gluon] (V)  -- (O1);
     \end{tikzpicture}} &
      $\displaystyle{\frac{i\gs}{\sqrt{2}}\sigma_1}$ \\
     \hline
  \end{tabular}
\end{table}

\subsection{Dyson-Schwinger equation}

\begin{figure}[!htbp]
  \centering
  \subfloat[]{
     \begin{tikzpicture}[baseline, scale=0.8]
      \coordinate (I) at (-1.5, 0);
      \coordinate (O) at (1.5, 0);
      \coordinate (C) at (0, 0);

      \draw[fermion] (I) -- ($(C)-(1, 0)$); \draw[fermion] ($(C)+(1, 0)$) -- (O);
      \draw (C) circle[x radius=1, y radius=0.5] node{$D$};
     \end{tikzpicture}
     $\quad=\quad$
     \begin{tikzpicture}[baseline, scale=0.8]
      \coordinate (I) at (-0.8, 0);
      \coordinate (O) at (0.8, 0);

      \draw[dashed, fermion] (I) -- (O);
     \end{tikzpicture}
     $\;+\;$
     \begin{tikzpicture}[baseline, scale=0.8]
      \coordinate (I) at (-1, 0);
      \coordinate (O) at (1, 0);
      \coordinate (C) at (0, 0);

      \draw (C) node[circle, draw](C1){$\Sigma$};
      \draw[dashed, fermion] (I) -- (C1); \draw[dashed, fermion] (C1) -- (O);
     \end{tikzpicture}
     $\;+\;$
     \begin{tikzpicture}[baseline, scale=0.8]
      \coordinate (I) at (-1.7, 0);
      \coordinate (O) at (1.7, 0);
      \coordinate (C) at (-0.7, 0);
      \coordinate (D) at (0.7,  0);

      \draw (C) node[circle, draw](C1){$\Sigma$};
      \draw (D) node[circle, draw](D1){$\Sigma$};
      \draw[dashed, fermion] (I) -- (C1);
      \draw[dashed, fermion] (C1) -- (D1);
      \draw[dashed, fermion] (D1) -- (O);
     \end{tikzpicture}
     $\;+\;\cdots$
  } \\
  \subfloat[]{
     \begin{tikzpicture}[baseline, scale=0.8]
      \coordinate (I) at (-1, 0);
      \coordinate (O) at (1, 0);
      \coordinate (C) at (0, 0);

      \draw (C) node[circle, draw](C1){$\Sigma$};
      \draw[dashed, fermion] (I) -- (C1); \draw[dashed, fermion] (C1) -- (O);
    \end{tikzpicture}
     $\quad=\quad$
     \begin{tikzpicture}[baseline, scale=0.8]
      \coordinate (I) at (-2, 0);
      \coordinate (V1) at (-1.3, 0);
      \coordinate (C) at (0, 0);
      \coordinate (O) at (2, 0);

      \draw[dashed, fermion] (I) -- (V1);
      \draw[fermion] (V1) -- ($(C)-(1, 0)$);
      \draw[gluon] (V1) arc[start angle=180, delta angle=-180, radius=1.3] node(V2){};
      \draw (C) circle[x radius=1, y radius=0.5] node{$D$};
      \draw[fermion] ($(C)+(1, 0)$) -- (V2.center);
      \draw[dashed, fermion] (V2.center) -- (O);
     \end{tikzpicture}
     \label{fig:self:energy:large:Nc:sum:gluon}   }
  \caption{Dressed quark propagator in the large-$\Nc$ limit. }
  \label{fig:self:energy:large:Nc:sum}
\end{figure}
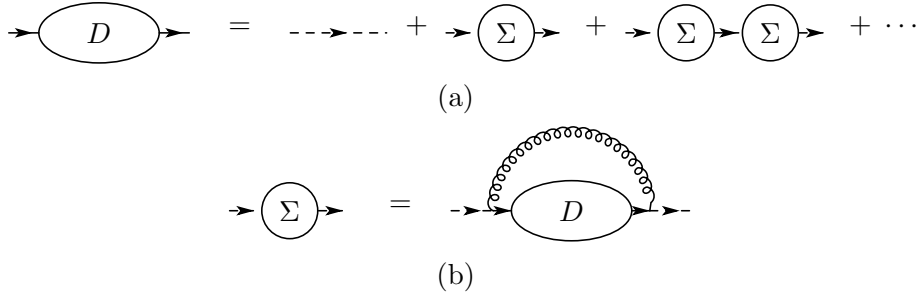

Let the contribution of the one-particle irreducible (1PI) diagrams to the
self-energy correction of quark propagator be denoted by $-i\Sigma\left(p\right)$.
In the large-$\Nc$ limit, the dressed quark propagator $D\left(p\right)$ is
determined by the rainbow diagrams (Fig. \ref{fig:self:energy:large:Nc:sum}).
It satisfies the following recursive equations:
\begin{subequations}
\begin{align}
  D\left(p\right) &= \frac{i}{p^0\sigma_1 - \omega_p^2 \mathcal{P}_{+} - \mathcal{P}_{-} -\Sigma\left(p\right)}, \\
  -i\Sigma\left(p\right) &= \frac{-i\gs^2\Nc}{2} \pint\frac{d^2 k}{\left(2\pi\right)^2}
  \frac{\sigma_1D\left(k\right)\sigma_1}{\left(p^z-k^z\right)^2},
    \label{eq:Sigma:by:D}
\end{align}
\end{subequations}
where the following expansion has been employed:
\begin{align}
  \frac{1}{X - Y} = X^{-1} + X^{-1}YX^{-1} + X^{-1}YX^{-1}YX^{-1} + \cdots.
\end{align}

Apparently $\Sigma\left(p\right)$ is independent of $p^0$. The most general form is
\begin{equation}
  \Sigma\left(p\right) = A\left(p^z\right) \mathcal{P}_{+} + B\left(p^z\right) \mathcal{P}_{-}.
\label{eq:Sigma:ex}
\end{equation}
The $\sigma_1$ term does not appear in \eqref{eq:Sigma:ex} because we take the
principal value of the $k^0$ integral in \eqref{eq:Sigma:by:D} at $k^0=\infty$,
i.e., we take the average of the contour closed at the upper half-plane and
the contour closed at the lower half-plane. The dressed quark propagator now
becomes
\begin{equation}
  D\left(p\right) =
    i\frac{p^0\sigma_1 + \left(1+B\right)\mathcal{P}_{+} +\left(\omega_p^2+A\right)\mathcal{P}_{-}}
      {\left(p^0\right)^2 - \left(1+B\right)\left(\omega_p^2+A\right)}.
\end{equation}
Introducing the following magic variables,
\begin{equation}
  E_p^2 \equiv \frac{\omega_p^2+A}{1+B},\quad
  F_p^2 \equiv \left(1+B\right)\left(\omega_p^2+A\right),
\label{eq:E:F}
\end{equation}
their advantage shows up immediately:
\begin{align}
  D\left(p\right) &=
    i\frac{p^0\sigma_1 + \frac{F_p}{E_p}\mathcal{P}_{+} + E_p F_p \mathcal{P}_{-}}
      {\left(p^0-F_p\right)\left(p^0+F_p\right)} \nn\\
  &= \frac{i}{2}\left(
    \frac{\sigma_1 + \frac{1}{E_p}\mathcal{P}_{+} + E_p\mathcal{P}_{-}}
      {p^0 - F_p + i\epsilon}
    + \frac{\sigma_1 - \frac{1}{E_p}\mathcal{P}_{+} - E_p\mathcal{P}_{-}}
      {p^0 + F_p - i\epsilon}
  \right),
\label{eq:D:ex}
\end{align}
where we have made assumptions that $\omega_p^2+A>0$, $1+B>0$.
Substituting \eqref{eq:D:ex} into \eqref{eq:Sigma:by:D}, we find
\begin{equation}
  A\left(p^z\right) \mathcal{P}_{+} + B\left(p^z\right) \mathcal{P}_{-} = \frac{\lambda}{2}\pint dk^z
    \frac{E_k\mathcal{P}_{+} + \frac{1}{E_k}\mathcal{P}_{-}}{\left(p^z-k^z\right)^2}.
\end{equation}
Matching the coefficients of $\mathcal{P}$ and utilizing \eqref{eq:E:F}, we find the
equations for $E_p$ and $F_p$:
\begin{align}
  F_p &= \frac{1}{E_p} \left(
    \omega_p^2 + \frac{\lambda}{2}\pint dk^z \frac{E_k}{\left(p^z-k^z\right)^2}
  \right)
  = E_p \left(
    1 + \frac{\lambda}{2}\pint dk^z \frac{1}{\left(p^z-k^z\right)^2}\frac{1}{E_k}
  \right).
\label{eq:E:F:sln}
\end{align}
(From \eqref{eq:E:F:sln} one can identify $E_p$ with $E_p$ in the Hamiltonian approach
and $F_p$ with $\Pi^+\left(p\right)$.)

Explicit calculation shows that the numerators of \eqref{eq:D:ex} can be
decomposed into outer products of two vectors:
\begin{subequations}
\begin{align}
  \frac{\sigma_1 + \frac{1}{E_p}\mathcal{P}_{+} + E_p\mathcal{P}_{-}}{2}
    &= \xi\left(p\right) \tilde{\xi}\left(p\right)\sigma_1, \\
  \frac{\sigma_1 - \frac{1}{E_p}\mathcal{P}_{+} - E_p\mathcal{P}_{-}}{2}
    &= \eta\left(p\right) \tilde{\eta}\left(p\right)\sigma_1,
\end{align}
\end{subequations}
where
\begin{subequations}
\begin{align}
  \xi\left(p\right) = \frac{1}{\sqrt{2}}\left(1, E_p\right)^\T,&\quad
  \tilde{\xi}\left(p\right) = \frac{1}{\sqrt{2}}\left(1, 1/E_p\right), \\
  \eta\left(p\right) = \frac{1}{\sqrt{2}}\left(1, -E_p\right)^\T,&\quad
  \tilde{\eta}\left(p\right) = \frac{1}{\sqrt{2}}\left(1, -1/E_p\right).
\end{align}
\end{subequations}
The dressed quark propagator can then be written as
\begin{align}
  D\left(p\right) &=
    \frac{i\xi\left(p\right)\tilde{\xi}\left(p\right)\sigma_1}
      {p^0 - F_p + i\epsilon}
    + \frac{i\eta\left(p\right)\tilde{\eta}\left(p\right)\sigma_1}
      {p^0 + F_p - i\epsilon}.
\label{eq:D:simp}
\end{align}

\subsection{Bethe-Salpeter equation}

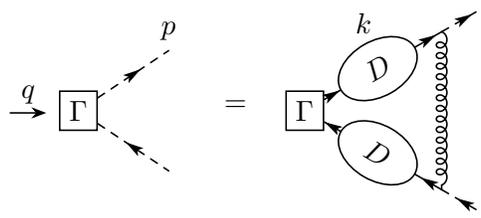
\begin{figure}[!htbp]
  \centering
  \begin{tikzpicture}[baseline, scale=0.8]
    \coordinate (V1) at (1.5, 1);
     \coordinate (V2) at (1.5, -1);

     \node[draw] (O) at (0, 0) {$\Gamma$};
     \draw[dashed, fermion] (O) -- (V1) node[above]{$p$};
     \draw[dashed, antifermion] (O) -- (V2);

     \draw (O.west) node[left, xshift=-5, yshift=6] (LB) {$q$};
    \draw[-Stealth] (LB.center) ++(-0.3, -0.3) -- +(0.6, 0);
  \end{tikzpicture}
  $\quad=\quad$
  \begin{tikzpicture}[baseline, scale=0.8]
     \coordinate (O) at (0, 0);
    \coordinate (V1) at (30:3.5);
     \coordinate (V2) at (-30:3.5);
     \coordinate (I1) at ($(O)!0.75!(V1)$);
     \coordinate (I2) at ($(O)!0.75!(V2)$);

     \node[draw] (ON) at (O) {$\Gamma$};
     \node[ellipse, rotate=30, draw] (I3) at ($(O)!0.4!(V1)$) {$\; D\;$};
     \node[ellipse, rotate=-30, draw] (I4) at ($(O)!0.4!(V2)$) {$\; D\;$};
     \draw[fermion] (ON) -- (I3);
     \draw[antifermion] (ON) -- (I4);
     \draw[fermion] (I3) -- (I1);
     \draw[antifermion] (I4) -- (I2);
     \draw[gluon] (I1) -- (I2);
     \draw[dashed, fermion] (I1) -- (V1);
     \draw[dashed, antifermion] (I2) -- (V2);

    \draw (I3.north) node[above]{$k$};
  \end{tikzpicture}

  \caption{Meson-$q\bar{q}$ vertex in the large-$\Nc$ limit. }
  \label{fig:meson-qqbar}
\end{figure}

In the large-$\Nc$ limit, The meson-$q\bar{q}$ vertex $\Gamma\left(p;q\right)$
satisfies the following Bethe-Salpeter equation (Fig. \ref{fig:meson-qqbar}):
\begin{equation}
  \Gamma\left(p; q\right) = \frac{-i\gs^2\Nc}{2}\pint\frac{d^2 k}{\left(2\pi\right)^2}
    \frac{1}{\left(p^z - k^z\right)^2}
    \sigma_1D\left(k\right)\Gamma\left(k; q\right)D\left(k-q\right)\sigma_1.
\label{eq:Gamma:rec}
\end{equation}

$\Gamma\left(p;q\right)$ is independent of $p^0$, so we can perform the $k^0$
integration:
\begin{equation}
  \mathcal{I}\left(k; q\right) \equiv \int\frac{dk^0}{2\pi}
    D\left(k\right)\Gamma\left(k; q\right) D\left(k-q\right).
\end{equation}
The interpretation of $\mathcal{I}\left(k; q\right)$ at the matrix-element level is
\begin{equation}
  \mathcal{I}\left(k; q\right) = \int dz\;e^{-ik^z z}
    \bra{\Omega} \Phi\left(0, z\right) \Phi^\dagger\left(0, 0\right)\ket{M_n\left(q\right)},
\end{equation}
where $\Phi\left(x\right) = \left(\phi\left(x\right), i\pi\left(x\right)\right)^\T$
and $M_n$ is the $n$-th excitation of the bound states.
Among the four projections of $D\left(k\right)$ and $D\left(k-q\right)$, only two
of them give nonzero residues. We arrive at
\begin{align}
  \mathcal{I}\left(k;q\right) &=
    \frac{
      \xi\left(k\right)\left[
        -i \tilde{\xi}\left(k\right)\sigma_1\Gamma\left(k;q\right)\eta\left(k-q\right)
      \right]\tilde{\eta}\left(k-q\right)\sigma_1
    }{q^0_n - F\left(k-q\right)-F\left(k\right) + i\epsilon} \nn\\
    &\qquad + \frac{
      \eta\left(k\right)\left[
        -i \tilde{\eta}\left(k\right)\sigma_1\Gamma\left(k;q\right)\xi\left(k-q\right)
      \right]\tilde{\xi}\left(k-q\right)\sigma_1
    }{- q^0_n - F\left(k-q\right)-F\left(k\right) + i\epsilon} \nn\\
  &\equiv -\chi^+_n\left(k^z;q^z\right) \sqrt{\frac{E_{k-q}}{E_k}}
      \xi\left(k\right)\tilde{\eta}\left(k-q\right)\sigma_1 \nn\\
    &\qquad - \chi^-_n\left(k^z;q^z\right) \sqrt{\frac{E_{k-q}}{E_k}}
      \eta\left(k\right)\tilde{\xi}\left(k-q\right)\sigma_1.
\label{eq:I:ex}
\end{align}
From now on we drop the superscript $z$ in spacial momenta for simplicity.
Substituting \eqref{eq:Gamma:rec} into \eqref{eq:I:ex}, we get
\begin{subequations}
\begin{align}
  &\left[q^0_n - F\left(p-q\right)-F\left(p\right)\right]\chi^+_n\left(p;q\right)
    = -\lambda \pint \frac{d k}{\left(p-k\right)^2} \sqrt{\frac{E_pE_{k-q}}{E_kE_{p-q}}} \nn\\
  &\qquad\quad \times\Big[
    \tilde{\xi}\left(p\right)\xi\left(k\right)\tilde{\eta}\left(k-q\right)\eta\left(p-q\right)
      \chi^+_n\left(k;q\right) \nn\\
  &\qquad\quad + \tilde{\xi}\left(p\right)\eta\left(k\right)\tilde{\xi}\left(k-q\right)\eta\left(p-q\right)
      \chi^-_n\left(k;q\right)
  \Big], \\
  &\left[-q^0_n - F\left(p-q\right)-F\left(p\right)\right]\chi^-_n\left(p;q\right)
    = -\lambda \pint \frac{d k}{\left(p-k\right)^2} \sqrt{\frac{E_pE_{k-q}}{E_kE_{p-q}}} \nn\\
  &\qquad\quad \times\Big[
    \tilde{\eta}\left(p\right)\eta\left(k\right)\tilde{\xi}\left(k-q\right)\xi\left(p-q\right)
      \chi^-_n\left(k;q\right) \nn\\
  &\qquad\quad + \tilde{\eta}\left(p\right)\xi\left(k\right)\tilde{\eta}\left(k-q\right)\xi\left(p-q\right)
      \chi^+_n\left(k;q\right)
  \Big].
\end{align}
\end{subequations}
Using the fact that
\begin{subequations}
\begin{align}
  \tilde{\xi}\left(p\right)\xi\left(k\right) =
  \tilde{\eta}\left(p\right)\eta\left(k\right) &=
    \frac{E_p + E_k}{2E_p}, \\
  \tilde{\xi}\left(p\right)\eta\left(k\right) =
  \tilde{\eta}\left(p\right)\xi\left(k\right) &=
    \frac{E_p - E_k}{2E_p},
\end{align}
\end{subequations}
and defining
\begin{subequations}
\begin{align}
  f_\pm\left(p, k\right) &\equiv \frac{E_k \pm E_p}{\sqrt{E_pE_k}}, \\
  S_\pm\left(p, k; q\right) &\equiv f_\pm\left(p,k\right)f_\pm\left(p-q,k-q\right),
\end{align}
\end{subequations}
the bound state equations can be written as
\begin{subequations}
  \begin{align}
    &\left[q^0_n - F\left(p-q\right)-F\left(p\right)\right]\chi^+_n\left(p;q\right)
      = -\frac{\lambda}{4} \pint \frac{d k}{\left(p-k\right)^2} \nn\\
    &\qquad\quad \times\left[
        S_+\left(p, k; q\right) \chi^+_n\left(k;q\right)
        - S_-\left(p, k; q\right) \chi^-_n\left(k;q\right)
    \right], \\
    &\left[-q^0_n - F\left(p-q\right)-F\left(p\right)\right]\chi^-_n\left(p;q\right)
      = -\frac{\lambda}{4} \pint \frac{d k}{\left(p-k\right)^2} \nn\\
    &\qquad\quad \times\left[
      S_+\left(p, k; q\right) \chi^-_n\left(k;q\right)
      - S_-\left(p, k; q\right) \chi^+_n\left(k;q\right)
    \right].
  \end{align}
  \end{subequations}
These equations are equivalent to \eqref{eq:scalar:ET:WE} with $F \equiv \Pi^+$.

\section{Connection of the mass renormalization schemes between equal-time and light-front quantization for scalar ${\rm QCD}_2$}
\label{appendix:two:renorm:mass}

In this appendix, we discuss some subtle issues about the mass renormalization in the scalar ${\rm QCD}_2$, in both equal-time and
light-front quantization.
We start from the equal-time quantization, and examine the asymptotic behavior of $E_k$ and $\Pi^+$ in the
large momentum limit.

The integrals appearing in $\Pi^\pm$ defined in \eqref{Pi:pm:f:pm:def} as $k\to\infty$ can be
calculated with the method of regions \cite{Beneke:1997zp}.
There are two regions of the integration variable $k_1$:
1) $m_{\tilde{r}} \sim \left|k_1\right| \ll k$;
2) $m_{\tilde{r}} \ll \left|k_1\right| \sim k$.
We divide the integration regions, expand the integrand according
to the different scaling of each region:
\begin{align}
  \pint dk_{1}\left(
    \frac{\frac{E_{k_1}}{E_k}\pm\frac{E_k}{E_{k_1}}}{(k+k_1)^{2}}
    -\frac{E_{k_1}}{E_k}\frac{1}{k_1^2}
  \right)
  &\approx \frac{1}{k}\pint_{\left|k_1\right|<\Lambda} dk_{1}
  \left(\pm\frac{1}{E_{k_1}}-\frac{E_{k_1}}{k_1^2}\right) \nn\\
  &\quad+ \pint_{\left|k_1\right|>\Lambda} dk_1\left(
    \frac{\frac{\left|k_1\right|}{k}\pm\frac{k}{\left|k_1\right|}}{\left(k+k_1\right)^2}
    -\frac{1}{k\left|k_1\right|}
  \right),
\end{align}
where $m_{\tilde{r}}\ll\Lambda\ll k$. For $\Pi^+$, the bounds of each region can be
extended to the whole momentum space since the asymptotic $\Lambda$ dependence
of each region cancels each other:
\begin{align}
  \pint dk_{1}\left(
    \frac{\frac{E_{k_1}}{E_k}+\frac{E_k}{E_{k_1}}}{(k+k_1)^{2}}
    -\frac{E_{k_1}}{E_k}\frac{1}{k_1^2}
  \right)
  &\approx \frac{1}{k}\left[
    \pint dk_{1} \left(\frac{1}{E_{k_1}}-\frac{E_{k_1}}{k_1^2}\right) - 4
  \right].
\end{align}
For $\Pi^-$, the integral is divergent for each region, but the divergences cancel each other in the sum of two  regions:
\begin{align}
  \pint dk_{1}\left(
    \frac{\frac{E_{k_1}}{E_k}-\frac{E_k}{E_{k_1}}}{(k+k_1)^{2}}
    -\frac{E_{k_1}}{E_k}\frac{1}{k_1^2}
  \right)
  &\approx \lim_{\Lambda/k_1\to\infty} \frac{1}{k} \left[
    \pint_{-\Lambda}^{\Lambda} dk_1 \left(-\frac{1}{E_{k_1}}-\frac{E_{k_1}}{k_1^2}\right) - 4 \ln\frac{k}{\Lambda}
  \right].
\end{align}
Define $k_0$ as
\beq
  4\ln k_{0} \equiv \lim_{\Lambda/k_1\to\infty} \pint_{-\Lambda}^{\Lambda} dk_1
    \left(-\frac{1}{E_{k_1}}-\frac{E_{k_1}}{k_1^2}\right)
    + 4\ln\Lambda.
\eeq
The asymptotic behavior of $E_{k}$ can be derived from
\eqref{eq:scalar:mass_gap:renorm} as
\begin{align}
  E_k &\approx k+\frac{1}{k}\left(\frac{m_{\tilde{r}}^{2}}{2}
    -\lambda\ln\frac{k}{k_0}\right).
\label{eq:E_k:asym}
\end{align}
The value of $E_k$ still can not be determined solely from \eqref{eq:E_k:asym}
since $k_0$ depends recursively on the functional form of $E_{k_1}$.
Fortunately, however, when substituting \eqref{eq:E_k:asym} into $\Pi^+$, the $k_0$
dependence cancels and we arrive at
\begin{align}
  \Pi^+\left(k\right) &\approx k+\frac{m_{\tilde{r}}^{2}}{2k}
  +\frac{\lambda}{4k}\left[\pint dk_1
    \left(\frac{1}{E_{k_1}}-\frac{E_{k_1}}{k_1^2}\right)-4\right],
\end{align}
By relabeling the momenta $k=xP$ in \eqref{eq:scalar:ET:WE} and taking the large $P$ limit, one
finds that \eqref{eq:scalar:ET:WE} approaches \eqref{eq:scalar:LC:WE:renorm}.
The relation between the renormalized masses $m_r$ and $m_{\tilde{r}}$ can be determined by
matching the two equations:
\begin{align}
  m_{\tilde{r}}^2 + \frac{\lambda}{2} \pint dk_1
    \left(\frac{1}{E_{k_1}}-\frac{E_{k_1}}{k_1^2}\right) &= m_r^2.
\end{align}
It is important to note that the relation between the equal-time renormalized mass and the light-front
renormalized mass is scheme-dependent. Had we instead chosen another equal-time renormalized
mass
\beq
  m_{\hat{r}}^{2} = m^2+\frac{\lambda}{2}\int dk_1\frac{1}{E_{k_1}},
\eeq
we would have found that $m_{\hat{r}} = m_r$.

\end{document}